\documentclass[reprint,superscriptaddress,amsfonts,amsmath,amssymb,aps,pre,floatfix,10pt]{revtex4-1}

\usepackage{graphicx}% Include figure files
\usepackage{dcolumn}% Align table columns on decimal point
\usepackage{bm}% bold math
\usepackage{xfrac}
\usepackage[normalem]{ulem}

 \usepackage{xcolor}
 
 \usepackage{tikz}

\newcommand\dd{\mathrm{d}}
\newcommand\pp{\partial}
\newcommand\tr{\mathrm{tr}}

\newcommand\Qvec{\mathbf{Q}}

\newcommand\x{\bm{x}}

\newcommand\nvec{{\bf n}}
\newcommand\mvec{{\bf m}}
\newcommand\n{{\bf n}}

\begin{document}

%\preprint{APS/123-QED}

\title{Dynamic tuning of the director field in liquid crystal shells %director field 
using block copolymers}% Force line breaks with \\

 \author{JungHyun Noh}
\affiliation{Department of Physics and Materials Science, University of Luxembourg}
 %\altaffiliation{Also at Cornell University.}%Lines break automatically or 
 
\author{Yiwei Wang}%
\affiliation{Department of Applied Mathematics, Illinois Institute of Technology, Chicago, IL 60616, USA}%

\author{Hsin-Ling Liang}
\affiliation{Institute of Organic Chemistry, Johannes Gutenberg University Mainz, Germany}
 %\altaffiliation{Also at Cambridge University.}%Lines break automatically or can be forced with \\
 
 \author{Venkata Subba Rao Jampani}%
\affiliation{Department of Physics and Materials Science, University of Luxembourg}%

 \author{Apala Majumdar} \email[Email: ]{apala.majumdar@strath.ac.uk}%
\affiliation{Department of Mathematics and Statistics, University of Strathclyde, Glasgow, United Kingdom}%

 \author{Jan P.F. Lagerwall} \email[Email: ]{jan.lagerwall@lcsoftmatter.com}%
\affiliation{Department of Physics and Materials Science, University of Luxembourg}%

%\email{jan.lagerwall@lcsoftmatter.com}
%\homepage{http://www.lcsoftmatter.com}

\date{\today}% It is always \today, today,

\begin{abstract}

When an orientationally ordered system, like a nematic liquid crystal (LC), is confined on a self-closing spherical shell, topological constraints arise with intriguing consequences that depend critically on how the LC is aligned in the shell. We demonstrate reversible dynamic tuning of the alignment, and thereby the topology, of nematic LC shells stabilized by the nonionic amphiphilic block copolymer Pluronic F127. Deep in the nematic phase, the director (the average molecule orientation) is tangential to the interface, but upon approaching the temperature $T_{NI}$ of the nematic--isotropic transition, the director realigns to normal. We link this to a delicate interplay between an interfacial tension that is nearly independent of director orientation, and the configuration-dependent elastic deformation energy of an LC confined in a shell. The process is primarily triggered by the heating-induced reduction of the nematic order parameter, hence realignment temperatures differ by several tens of degrees between LCs with high and low $T_{NI}$, respectively. The temperature of realignment is always lower on the positive-curved shell outside than at the negative-curved inside, yielding a complex topological reconfiguration on heating. Complementing experimental investigations with mathematical modeling and computer simulations, we identify and investigate three different trajectories, distinguished by their configurations of topological defects in the initial tangential-aligned shell. Our results uncover a new aspect of the complex response of LCs to curved confinement, demonstrating that the order of the LC itself can influence the alignment and thereby the topology of the system. They also reveal the potential of amphiphilic block copolymer stabilizers for enabling continuous tunability of LC shell configuration, opening doors for in-depth studies of topological dynamics as well as novel applications in, e.g., sensing and programmed soft actuators. 
\end{abstract}

\maketitle

%\tableofcontents

\section{\label{sec:level1}Introduction}
Double emulsions of water in liquid crystal (LC) in water, also known as LC shells, have over the last decade acquired a status as a prolific experimental platform for studying confinement effects in soft matter physics \cite{urbanski2017liquid,lagerwall2012new,lopez2011drops}, in particular concerning topological defects and their interactions on curved spaces. While the initial work, experimental and theoretical, was restricted to nematic shells (orientational order  only; molecules aligning along the director \textbf{n}) \cite{nelson2002toward,fernandez2007novel,skacej2008controlling,lopez2009topological,kralj2011curvature,lopez2011frustrated,napoli2012surface,napoli2012extrinsic}, later efforts have focused also on smectic shells of SmA- \cite{liang2011liquid,liang2011nematic,lopez2011nematic,liang2012towards,lopez2012smectic,liang2013tuning,manyuhina2015thick,akita2017room} and SmC-type \cite{liang2013tuning} as well as cholesteric \cite{uchida2013controlled,geng2018through,geng2017elucidating,tran2017change,darmon2016waltzing,darmon2016topological,geng2016high} shells. %These studies revealed intriguing frustration effects related to the development of the discrete layers of smectics and the helical director modulation of cholesterics, respectively, in self-closing spherical confinement subject to varying boundary conditions. 
Thanks to advances in providing long-term stability through polymerization/polymer stabilization of LC shells \cite{lee2017structural,heo2017smart,noh2016taming,geng2016high}, they are also emerging as a realistic basis for innovative applications, for instance in photonics and photonics-derived use cases \cite{iwai2020shrinkage,myung2019optical,schwartz2018cholesteric,lee2017structural,lenzini2017security,kim2017photonic,kang2017amplified,geng2016high,uchida2013controlled}, sensing \cite{sharma2019realignment,heo2017smart,jang2017ph} or unconventional soft actuators \cite{jampani2019liquid,jampani2018micrometer,fleischmann2012one}.

For research as well as applications, it is imperative to control the director field configuration within the shell, a requirement typically fulfilled by choosing interface stabilizers that promote the desired orientation of \textbf{n} at each boundary \cite{Sharma2018influence,noh2016influence}. Surprisingly little efforts have been devoted to exploring the vast parameter space of potential stabilizer molecules. One stabilizer, $\sim85$\% hydrolyzed polyvinylalcohol (PVA), dominates entirely in work where tangential alignment (often also called planar alignment) is desired \cite{fernandez2007novel}, and for normal (often called homeotropic) alignment the majority of studies use the anionic surfactant sodium dodecylsulfate (SDS) \cite{noh2016influence,lopez2009topological}. A recent systematic screening of cat- and anionic low molar mass surfactants revealed, however, that their impact is more subtle than expected \cite{Sharma2018influence}; depending on surfactant length and concentration, a fully tangential ('T'), a hybrid ('H', tangential on one side, normal on the other) or a fully normal ('N') configuration can be achieved. It is also possible that the stable alignment at a boundary is tilted, between tangential and normal, as has been demonstrated for lecithin-stabilized nematic droplets in glycerol \cite{Volovik.1983}. Our understanding of how stabilizers affect the director field configuration in LC shells is thus far from complete.

Even with well-defined uniform boundary conditions, there may be significant variability in the director field and distribution of topological defects in the shell. This is particularly the case for shells in a T configuration, a situation that has been extensively investigated, experimentally \cite{noh2016taming,sevc2012defect,liang2011nematic,lopez2011nematic,lopez2011frustrated,fernandez2007novel} as well as theoretically and by computer simulation \cite{nelson2002toward,koning2016spherical,koning2013bivalent,napoli2012surface,napoli2012extrinsic,sevc2012defect,lopez2011frustrated,kralj2011curvature,skacej2008controlling}. The Poincar\'e--Hopf theorem dictates that the inside as well as the outside of a T shell must have a total topological defect charge of +2 \cite{urbanski2017liquid,lopez2011drops}. This requirement can be fulfilled in three ways, all of which have been found experimentally and in simulation: 
\begin{enumerate}
\item Four +\sfrac{1}{2} disclination lines connect the shell inside and outside. We will refer to this tangential configuration, which is the most common, as 'T1', alternatively the more descriptive label $\mathbf{4(+\sfrac{1}{2})}$. 
\item Two +\sfrac{1}{2} disclination lines connect the shell inside and outside, the remaining +1 topological charge at each interface being realized by a point defect. We call this configuration 'T2', with the descriptive label $\mathbf{2(+\sfrac{1}{2}),+1}$. 
\item The inside as well as the outside each have two +1 point defects. We label this tangential configuration 'T3', alternatively $\mathbf{2(+1)}$. 
\end{enumerate}

The H configuration has so far been investigated in detail only experimentally \cite{liang2012towards,liang2013tuning,noh2016influence}, revealing that the observable stable hybrid configuration always has two antipodal +1 point defects on the tangentially aligned side of the shell, one at the thinnest and one at the thickest point. A shell in the N configuration is defect-free, as its radial director field has only a virtual bulk defect at the center of the shell, which is occupied by the isotropic liquid core.

\begin{figure}
\centering
 \includegraphics[width=8.5cm]{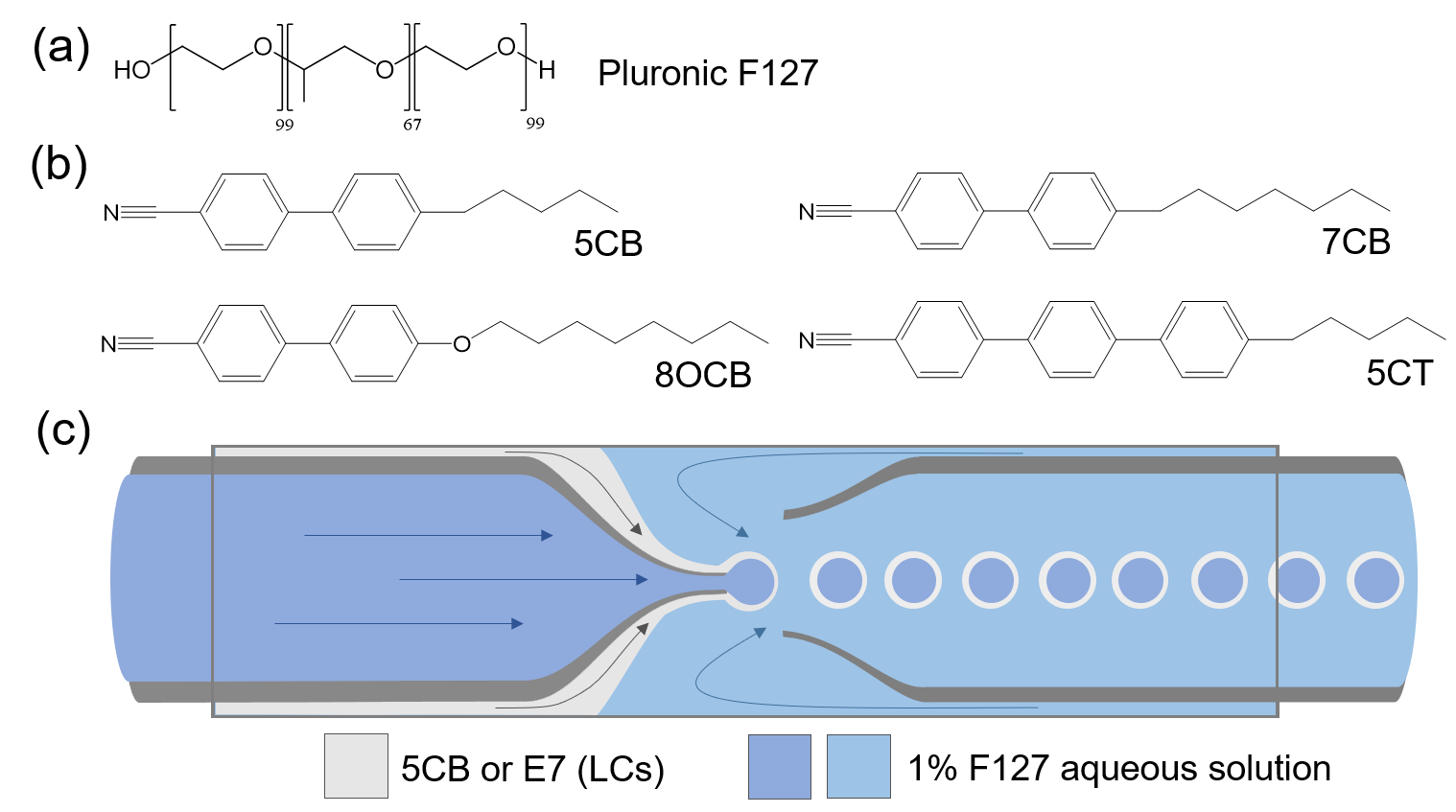}
 \caption{Molecular structures of (a) Pluronic F127 block copolymer and (b) mesogens; 5CB, 7CB, 8OCB and 5CT. The mixture of the four components yields E7 (mixing ratio in Section~\ref{exp}). (c) Schematic illustration of the production of LC shells using a coaxial microfluidic device.}
 \label{fgr:schematics}
\end{figure}

While most LC shells have permanent configuration once produced, it is very powerful to be able to dynamically tune the director field after production, for scientific experiments and applications. Such realignment has been achieved twice for shells, first via an exchange of stabilizer in the outer phase \cite{lopez2009topological}, then using a custom-synthesized light-switchable additive \cite{noh2018sub}. For droplets, Volovik and Lavrentovich achieved realignment of the single outer interface simply by heating lecithin-stabilized nematic droplets dispersed in glycerol throughout the nematic temperature range \cite{Volovik.1983}. Deep in the nematic phase, the droplets were normally aligned, while they adopted tangential alignment near the clearing point. The change was mediated via a continuous increase of director tilt with respect to the interface upon heating. 

Here we demonstrate that temperature-driven reversible dynamic tuning of the configuration is possible also with nematic shells in water, using the commercially available amphiphilic block copolymer Pluronic F127 (Fig.~\ref{fgr:schematics}a) for stabilizing the interfaces. F127 consists of two hydrophilic polyethyleneoxide (PEO) blocks separated by a hydrophobic polypropyleneoxide (PPO) block, with the overall composition PEO$_{99}$-PPO$_{67}$-PEO$_{99}$. In preliminary experiments \cite{liang2011liquid,liang2013tuning} we noticed the remarkable ability of F127-stabilized shells to change alignment with temperature $T$: at room temperature the shells are in a T configuration, but on heating they switch first to H and then to N configuration near the nematic--isotropic transition (clearing) at $T=T_{NI}$. These early studies were carried out on shells of 4-Octyl-4'-cyanobiphenyl (8CB), with a phase sequence Cr. 21.5$^\circ$C SmA 33$^\circ$C N 40.5$^\circ$C iso. We thus speculated that the configuration changes may have been related to the development/disappearance of smectic order. 

In the current study we refute this initial hypothesis, by presenting detailed investigations of shells of two LC materials (Fig.~\ref{fgr:schematics}b) without smectic phases, 4-cyano-4'-pentyl biphenyl (5CB) and the mixture E7. We find the same sequence of reversible configuration changes, T${}^{\phantom{1}\rightarrow}_{\textrm{heat}}$H${}^{\phantom{1}\rightarrow}_{\textrm{heat}}$N${}^{\phantom{1}\rightarrow}_{\textrm{cool}}$H${}^{\phantom{1}\rightarrow}_{\textrm{cool}}$T, on heating and subsequent cooling, showing that smectic order is not required. It is primarily the variation of orientational order with temperature that drives the changes. Significantly, the temperatures at which configurational changes take place are very different for the two LCs, a consequence of the higher clearing point of E7 ($T_{NI}\approx59^\circ$C) than of 5CB ($T_{NI}=35.5^\circ$C). We complement the experiments with theoretical modelling and numerical results, providing new insights into nematic equilibria in hybrid shells, twisted director profiles and dynamical pathways from tangential to hybrid, that mimic experiments.

%F127 has been studied extensively as liposome stabilizer to improve drug delivery efficiency.\cite{Chandaroy2002,Zhang2005e} This polymer changes its characteristics with temperature as both the PPO and PEO moieties dehydrate on heating, the former more rapidly than the latter. This gives F127 solutions the peculiar property of a critical micelle temperature (CMT), below which no micelles form.\cite{Zhang2005e}

\section{Experiment and simulation details}
\subsection{Experimental shell production and characterization}\label{exp}
LC shells are fabricated using a glass capillary-based microfluidic emulsification device, see Fig.~\ref{fgr:schematics}c, all capillaries from WPI Instruments. Two cylindrical capillaries (1 mm outer diameter) are coaxially fitted into a square capillary (inner dimension 1 mm), the former tapered at one end and carefully aligned with nozzles facing each other. These serve as injection and collection tube, respectively. Unless otherwise stated, a 1~wt.\% aqueous solution of F127 serves as inner as well as outer phase. The inner phase is pumped through the injection tube with the LC forming the middle fluid, flowing in the same direction through the voids between injection tube and square capillary. It thus coats the inner fluid at the orifice. The outer phase is flown in the opposite direction through the voids between collection tube and square capillary, flow-focusing the LC--inner phase co-flow into the collection tube, where the compound jet breaks into discrete LC shells. The injection tube is hydrophobically treated to prevent wetting by the inner aqueous phase at the tip region. A temperature controlled heating cavity houses the device, so LC shells can be produced in the isotropic phase where the flow is the easiest. The diameter of the produced shells is in the range 100--120~$\mu$m and the average thickness 3--6~$\mu$m.

\textbf{Liquid crystals:} We make shells from 5CB (Cr. 24 N 35.5 Iso./$^\circ$C; the nematic phase is easily supercooled to $\sim20^\circ$C) as well as from the four-component mixture E7 (nematic on cooling to -60$^\circ$C, clearing range on heating about 58.5--60$^\circ$C), composed of the four cyanobiphenyl-based mesogens 5CB (47 mol\%), 7CB (4-cyano-4'-heptyl biphenyl, 25 mol\%), 8OCB (4-cyano-4'-octyloxy biphenyl, 18 mol\%) and 5CT (4-cyano-4'-pentyl terphenyl, 10 mol\%). All structures are shown in Fig.~\ref{fgr:schematics}.

\textbf{POM and FCPM characterization:} Shell suspensions are filled into rectangular capillaries (200~$\mu$m thickness) and observed in a polarizing optical microscope (POM, Olympus BX51) equipped with a hot stage. For fluorescence confocal polarization microscopy (FCPM, Nikon A1R+) to map out the LC director field \cite{smalyukh2001three, noh2018sub} a small amount (0.1$\%$) of dichroic dye (BTBP: N,N'-bis(2,5-di-tert-butylphenyl)- 3,4,9,10-perylene-carboximide) is mixed into the LC before shell production. A capillary holding the shells is placed on a home-built resistive heating device, consisting of a 1~mm thick glass plate coated with transparent ITO (indium tin oxide) electrodes connected to a temperature-regulated power supply. The BTBP dye is excited by 488~nm laser light and the fluorescence emission is collected at 525~nm. The data are visualized in 3D using Nikon NIS elements imaging software.

\subsection{Theoretical and numerical methods}
In this section, we summarise our theoretical methods for simulating experimentally observable equilibrium nematic textures and their transient dynamics inside asymmetric shells. The shell is defined by 
$$\Omega = B(\mathbf{0}, R) \backslash B(\x_c, R_1) \subset \mathbb{R}^3,$$
where $B(\x_c, R_1) \subset B(\mathbf{0}, R)$. $B(\mathbf{0},R)$ denotes a three-dimensional ball of arbitrary radius $R$ centred at the origin, $B(\x_c, R_1)$ is a smaller ball of radius $R_1$ centred at the point $\x_c = (0,0,cR) \in B(0,1)$ for $c<1$. The shell has two boundaries, the outer boundary is $\partial B(\mathbf{0},R)$ whereas the inner boundary is $\partial B(\x_c, R_1)$. In what follows, we model shells with preferred tangential anchoring on $\partial B(\x_c, R_1)$ and with either tangential anchoring on $\partial B(\mathbf{0},R)$ (referred to as fully  tangential shells) or with normal anchoring on $\partial B(\mathbf{0},R)$ (referred to as hybrid shells).

We work within the powerful continuum Landau-de Gennes (LdG) theory for nematic LCs. The LdG theory describes the nematic phase by a macroscopic LdG $\mathbf{Q}$-tensor parameter, which is a symmetric traceless $3\times 3$ matrix that encodes information about the state of orientational order \cite{dg, majumdar2010}. The phase is said to be isotropic if $\mathbf{Q}=0$ with no orientational ordering, uniaxial nematic if $\mathbf{Q}$ has a single distinguished eigendirection, the usual director \textbf{n}, and biaxial nematic if $\mathbf{Q}$ has three distinct eigenvalues, rendering a primary and secondary director. A uniaxial $\mathbf{Q}$-tensor can be written as
$$ \mathbf{Q} = s \left(\nvec \otimes \nvec - \frac{\mathbf{I}}{3} \right)$$
where $s$ is a scalar order parameter that measures the degree of ordering about $\nvec$. A further metric of orientational ordering is the biaxiality parameter
$$\beta = 1 - 6 \frac{ (\tr \Qvec^3)^2}{ (\tr \Qvec^2)^3}.$$ One can show $0 < \beta \leq 1$ in the biaxial region, while $\beta = 0$ in the uniaxial region \cite{dg, wangmajumdarJMAA}. Biaxiality is particularly important for visualising and understanding the structure of nematic defect cores.

The LdG theory is a variational theory so that the experimentally observable configurations are modelled by either local or global minimizers of an appropriately defined LdG free energy \cite{dg, virga}. We work with the following well accepted form of the LdG free energy
\begin{equation}\label{LdG}
\mathcal{F}[\Qvec] = \int_{\Omega} f_{\rm b}(\Qvec) + f_{\rm el}(\Qvec, \nabla \Qvec) \dd V + \int_{\pp \Omega} f_{\rm s}(\Qvec) \dd S, 
\end{equation}
where $f_{\rm b}$ is the bulk energy density, $f_{\rm el}$ is the elastic energy density, and $f_{\rm s}$ is the surface energy density. The bulk energy density, $f_{\rm{b}}$, is given by
\begin{equation}
f_{\rm b} (\Qvec) =  \frac{A}{2} \tr(\Qvec^2) - \frac{B}{3} \tr(\Qvec^3) + \frac{C}{4} (\tr(\Qvec^2))^2,
\end{equation}
 where $A = a(T - T_{NI}^{*})$, with $T_{NI}^*$ the nematic supercooling limit such that the isotropic phase is unstable for $T< T_{NI}^*$, and $a, B$ and $C$ are material dependent parameters. %The bulk energy density, $f_{\rm b}$ favours ordered uniaxial nematic phases for $A<0$. 
 We can explicitly compute the minimizers of $f_{\rm b}$. For $A < \frac{B^2}{27 C}$, $f_{\rm b}$ attains its minimum on the set of uniaxial $\Qvec$-tensors defined by
 $$ \mathcal{N} = \left\{   \Qvec = s^+ \left(\mvec \otimes \mvec - \frac{\mathbf{I}}{3} \right)\right\}, $$
 where $s^+ = \frac{B + \sqrt{B^2 - 24 AC}}{4 C}$ is referred to as the ``optimal" scalar order parameter and $\mvec$ is an arbitrary unit-vector field.
For a given set of material constants, the isotropic-nematic phase transition takes place at $A = \frac{B^2}{27 C}$ \cite{dg, newtonmottram}. The elastic energy density is often given by
\begin{equation}
\begin{aligned}
f_{\rm el}(\Qvec, \nabla \Qvec) & = \frac{L_1}{2} {Q}_{ij,k} {Q}_{ij,k} + \frac{L_2}{2} {Q}_{ij,j}{ Q}_{ik,k} \\ 
        & + \frac{L_3}{2} {Q}_{ik,j}{ Q}_{ij,k} + \frac{L_4}{2}{Q}_{kl}{Q}_{ij,k} {Q}_{ij,l},\\
\end{aligned}
\end{equation}
where the $L_i$ ($i=1\dots4$) are experimentally measurable material elastic constants \cite{ballmajumdar2010}. %It is known that $f_{\rm el}$ is unbounded from below, so that energy minimization problems are ill-posed with $L_4 \neq 0$ \cite{ballmajumdar2010}
For relative simplicity and well-posedness, we take $L_3 = L_4=0$ in what follows (see \cite{ballmajumdar2010} for problems with $L_4 \neq 0$). The elastic constants can be correlated to certain specific material deformations e.g. splay, twist or bend. This can be more clearly seen if we compare $f_{\rm el}$ with the Oseen-Frank (OF) free energy density for uniaxial nematics, with $\mathbf{Q} = s\left(\mathbf{n}\otimes \mathbf{n} - \frac{\mathbf{I}}{3} \right)$ and constant $s$ \cite{newtonmottram}:
\begin{equation}
\begin{aligned}
W[\n] & = K_1( \nabla \cdot \n)^2 + K_2 (\n \cdot (\nabla \times \n))^2
                 + K_3 |\n \times (\nabla \times \n)|^2  \\
                 & + (K_2 + K_4) (\tr(\nabla \n)^2 - (\nabla \cdot \n)^2). \\
\end{aligned}
\end{equation}
Here $K_1$, $K_2$ and $K_3$ are the elastic constants that describes the energetic penalty of splay, twist and bend deformations, respectively. $K_4$ is a saddle-splay elastic constant usually associated with surface effects; it does not contribute to the Oseen-Frank energy of tangentially aligned nematic droplets and we do not include it in this manuscript \cite{dg, Williams}. Tran et al. compared simulations with and without saddle-splay constants for long-pitch cholesteric shells \cite{tran2017change}, concluding that the approximation $K_{4}=0$ introduces no significant errors. 

Whilst the OF theory is not as general as the LdG theory, we can make a correspondence between the elastic constants in the LdG theory and the OF theory as shown below, with $L_3 = L_4 =0$ \cite{newtonmottram}.% as shown below \cite{newtonmottram}
 %\begin{equation}\label{LiKi}
 %%\begin{aligned}
% & K_1 = 2 L_1 s^2 + L_2 s^2 + L_3 s^2 - \frac{2}{3}L_4 s^3, \\
% & K_2 = 2L_1 s^2 - \frac{2}{3}L_4 s^3, \\
% & K_3 = 2 L_1 s^2 + L_2 s^2 + L_3 s^2 + \frac{4}{3}L_4 s^3, \\
%& K_4 = L_3 s^2. \\
 %\end{aligned}
 %\end{equation}
%For relative simplicity and well-posedness, we set $L_3 = L_4=0$ so that
%or $L_3 = L_4=0$, we have
\begin{equation}
\begin{aligned}
& K_1 = (2 L_1  + L_2)s^2, \quad K_2 = 2 L_1 s^2, \\
& K_3 = (2 L_1  + L_2)s^2, \quad K_4 = 0, \\
\end{aligned}
\end{equation}
which implies $K_2 < K_1 = K_3$ for $L_2 > 0$ and
\begin{equation}
\frac{K_1}{K_2} = 1 + \frac{1}{2} \frac{L_2}{L_1}
\end{equation}
We define $\eta = L_2/L_1$ so that $K_1/K_2 = 1 + \eta/2$. In particular, increasing $\eta$ is equivalent to an increased energetic penalty for splay deformations, so that twisted configurations are more readily observable for large and positive $\eta$ (see \cite{Williams} for formal calculations of the energies of twisted configurations on tangentially aligned nematic droplets). The case $\eta= 0$ describes the one-constant approximation with $K_1=K_2 = K_3$ and $K_4=0$ \cite{dg}.

A further key modelling ingredient are the boundary conditions. In the case of hybrid shells, we impose normal anchoring on the outer shell by means of a Rapini-Papoular energy \cite{luo2012multistability}
\begin{equation}\label{fs}
{F}_{\rm s}  = \frac{W_0}{2} \int_{\partial B(\mathbf{0}, R)} ( Q_{ij}(\mathbf{x}) - Q_{ij}^{\rm s}(\mathbf{x}))^2 \dd S,
\end{equation}
where $W_0$ is the anchoring strength,
\begin{equation}
\Qvec^{\rm s}(\x) = s^{+} \left( \bm{\nu}(\x) \otimes \bm{\nu}(\x) - \frac{1}{3} \mathbf{I} \right), \quad \x \in \pp \Omega,
\end{equation}
and $\bm{\nu}$ is the surface normal of $\partial B(0, R)$. In the limit of $W_0 \to + \infty$, we recover strong normal anchoring since $F_{\rm s}$ is minimized by $\Qvec = \Qvec^{\rm s}$ for $W_0>0$. We do not consider $W_0<0$ in this manuscript but $W_0<0$ favours boundary alignments that are ``far" away from the idealised normal anchoring $\Qvec^{\rm s}$; this could be tangential anchoring, tilted anchoring etc. 

The tangential degenerate anchoring is imposed on the inner surface, $\partial B(\x_c, R_1)$ by the following surface energy  \cite{fournier2005modeling}
\begin{equation}\label{tangential_fs_1}
F_s =  \int_{\pp B(\x_c, R_1)} \frac{W_1}{2} \left( \widetilde{Q}_{ij} - \widetilde{Q}_{ij}^{\perp} \right)^2  \dd S,
\end{equation}
where $\widetilde{Q}_{ij} = Q_{ij} + \frac{1}{3}  s^{+} \delta_{ij}$,
\begin{equation}
\widetilde{Q}_{ij}^{\perp} = P_{ik} \widetilde{Q}_{kl}P_{lj}, \quad \mathbf{P} = \mathbf{I}  - \bm{\nu} \otimes \bm{\nu}.
\end{equation}
The surface energy in (\ref{tangential_fs_1}) favours the tangential projection of $\mathbf{Q}$ onto the surface of $\partial B(\x_c, R_1)$ without imposing specific directors i.e. it is minimised for $\widetilde{Q}_{ij} = \widetilde{Q}_{ij}^{\perp}$ for $W_1>0$ and $\widetilde{Q}_{ij}^{\perp}$ is the tangential projection of $\widetilde{Q}$ on $\partial B(\x_c, R_1)$. We note that $F_s$ in (\ref{tangential_fs_1}) could include tilted anchoring for moderate values of $W_1$ but we do not consider intermediate values in this manuscript. %The second term favours perfect uniaxial ordering on the surface with the optimal order parameter $s^+$  and hence, penalises biaxiality  \cite{tasinkevych2012}. %We take $W_2 = 0$ to reduce the number of model parameters.

It is important to non-dimensionalise the LdG free energy, so that we can study the dependence of the equilibria on key material characteristics e.g. elastic anisotropy, relative temperature, anchoring and geometrical size. We non-dimensionalize the free-energy (\ref{LdG}) by using $R$ as the characteristic length. The re-scaled domain becomes $\Omega = D_0 \backslash D_1$, 
\begin{equation}
D_0 = B(\mathbf{0}, 1), \quad D_1 =  B(\x_c, \rho),
\end{equation}
where $\rho = R_1 / R$, $\x_c = (0, 0, c)$ as before. The parameters $c$ and $\rho$ satisfy  $c > \rho > 0$ and $c + \rho < 1$. These parameters control the asymmetry and thickness of the shell. For example, $c=0$  describes a symmetric shell whereas $c>0$ describes a shell that this thinner towards the top. Smaller values of $c+ \rho$ describe thicker shells since the shell thickness is proportional to the quantity, $1-c-\rho$. 
%, which is illustrated in Fig. \ref{domain}(a). 
Hence, the rescaled boundary is :$\pp \Omega = \pp D_0 \cup \pp D_1$. Let $$\bar{\x} = \x/R, \quad \bar{\Qvec} = \sqrt{\frac{27C^2}{2B^2}}\Qvec, \quad  \bar{\mathcal{F}} = \frac{27^2C^3}{2B^4R^3}\mathcal{F},$$ 
Dropping all \emph{bars} for convenience, the dimensionless LdG free energy for a hybrid shell is given by
\begin{equation}\label{LdG_dimless}
  \begin{aligned}
    \mathcal{F}[\Qvec] & = \int_{\Omega} \frac{t}{2} \tr(\Qvec^2) - \sqrt{6} \tr(\Qvec^3) + \frac{1}{2} (\tr (\Qvec^2))^2 \\
    & + \frac{\xi_R^2}{2} ( Q_{ij,k}Q_{ij,k} + \eta  Q_{ij,j}Q_{ik,k}) \dd \mathbf{x} \\
    & + \int_{D_1} \frac{w_0}{2}  (Q_{ij}(\mathbf{x}) - Q_{ij}^{\rm s}(\mathbf{x}))^2  \dd A \\
    & +  \int_{D_2} \frac{w_1}{2} \left( \widetilde{Q}_{ij} - \widetilde{Q}_{ij}^{\perp} \right)^2   \dd A \\
  \end{aligned}
\end{equation}
where $t = \frac{27AC}{B^2}$ is a dimensionless temperature, $\xi_R  = \sqrt{\frac{27CL_1}{B^2R^2}}$ is related to the shell size, $\eta = \frac{L_2}{L_1}$ ($K_1/K_2 = 1 + \frac{1}{2} \eta$) is a measure of the elastic anisotropy. The nondimensionalized anchoring strength is given by $w_{i} = \frac{27 C W_i}{B^2 R}$ ($i = 0, 1$). In the simulation, we adopt the parameter values $A = -0.172 \times 10^{6}\  \mathrm{Jm}^{-3}, B = -2.12 \times 10^{6}\  \mathrm{Jm}^{-3}$,  $C = 1.73 \times 10^6\  \mathrm{Jm}^{-3}$, and $L_1 = 4 \times 10^{-11}\  \mathrm{Jm}^{-1}$, which are typical phenomenological values for 5CB at room temperature \cite{Muvsevivc2006two, ravnik2009landau}.   This motivates us to use $t = -1.79$ for most of the simulations, since  $t = -1.79$ corresponds to the room temperature for 5CB, whilst $t = 1$ is the nematic-isotropic transition temperature. The anchoring strength $W_0$ and $W_1$ are taken to be $10^{-2} {\rm J m}^{-2}$ to account for the strong anchoring \cite{ravnik2009landau}, if not stated differently.

% the equilibrium clearing temperature $T_{NI}$ (also referred to as $T_c$, the critical temperature), at which the  phases have identical free energies, corresponds to $t=1$. A metastable nematic phase can be superheated up to $t = 9/8$, where the local energy minimum for nematic order disappears, and a metastable isotropic phase can be supercooled to $t=0$, where the local minimum for isotropic disorder disappears.

The details of the numerical methods can be found in Appendix~\ref{ApxNum}. We use the spectral method to discretize the tensor order parameter $\Qvec$, that is expanding $\Qvec$ in terms of proper basis functions after the appropriate identification of a coordinate system \cite{wang2017topological}.  Neglecting the high-order terms in a truncated series, we express the free energy in terms of undetermined coefficients and use  standard optimization methods, to minimize the discrete free energy and compute local or global energy minimizers. For the dynamical transitions, we solve a gradient flow system for $\Qvec$, which are based on the principle that the free energy decreases with time till we settle into a local energy minimum.

Our numerics are limited to shells with diameter on the micron scale (1--5~$\mu$m), similar to other theoretical studies \cite{gharbi2013microparticles, sadati2017spherical} and are hence two orders of magnitude smaller than the shells in experiments (100--150~$\mu$m diameter). In some situations, this may lead to different results.

\section{Results}

\subsection{5CB shell textures on heating/cooling}\label{5cbshelldescript}
\subsubsection{Experimental observations}\label{5cbexp}
\begin{figure*}
\centering
 \includegraphics[width=17cm]{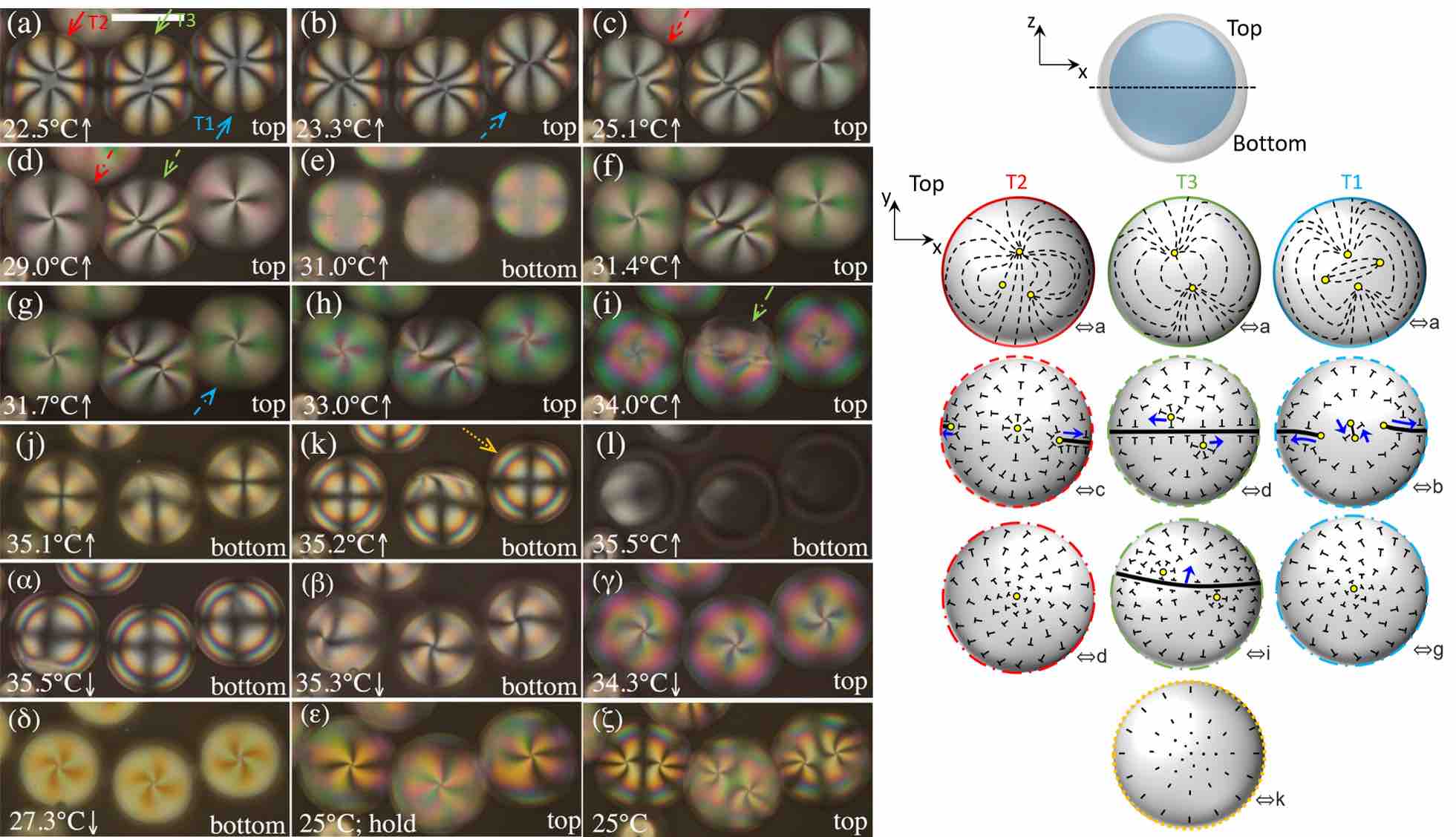}
 \caption{Freshly produced 5CB shells heated from room temperature to isotropic (a--l) and then cooled back ($\alpha$--$\epsilon$) at moderate cooling rate, holding at 25$^\circ$C at the end ($\epsilon-\zeta$). Initially (a), the three shells  have one each of the possible tangential configurations: T1 on the right, T2 on the left, and T3 in the middle. However, after the cooling experiment, all shells end up as T3. The shells are viewed along gravity, the texts "top" and "bottom" indicating the focal plane. Near room temperature they appear elliptical due to an optical artifact that is explained in Appendix~\ref{ApxPOM}. Scale bar: 100~$\mu$m. On the top right, a sketch of a shell from the side is drawn to illustrate the asymmetry, with thin top and thick bottom. No photo corresponds to this perspective. Below are sketches drawn in the same perspective as the photos, showing the top of each shell as viewed along gravity, of the director field for representative steps in the heating sequence. The color and character of the ring around each drawing matches those of the arrow pointing to the corresponding shell photo on the left. Localized defects are drawn in yellow and defect lines (explained in Section~\ref{E7-F127-shells}) as black lines. The blue arrows indicate direction of motion of a defect or line.}
 \label{fgr:5cbshells}
\end{figure*}

We heat 5CB shells surrounded by 1\% aqueous solutions of F127 from room temperature to slightly above $T_{NI}=35.5^\circ$C and then cool back. This experiment, as observed through the POM, is shown in Movie 1 in Ref.~\cite{supplement}, with representative snapshots on three adjacent shells in Fig.~\ref{fgr:5cbshells}. The shells start out pristine, i.e., they have gone through no temperature changes after the rapid cooling from isotropic to nematic at the end of the production process. The director configurations at room temperature (Fig.~\ref{fgr:5cbshells}a) correspond to tangential boundary condition on in- and outside (T configuration). The three selected shells represent all possible combinations of topological defects in a T configuration, as defined above:
\begin{enumerate}
    \item The T1 configuration [4(+\sfrac{1}{2}$^t$)] is seen in the right shell, highlighted by a blue arrow and labelled "T1" in Fig.~\ref{fgr:5cbshells}a.
    \item The T2 configuration [2(+\sfrac{1}{2}$^t$),+1$^t$] is seen in the left shell (red arrow), labelled "T2" in Fig.~\ref{fgr:5cbshells}a.
    \item The T3 configuration [2(+1$^t$)] appears in the middle shell (green arrow), labelled "T3" in Fig.~\ref{fgr:5cbshells}a.
\end{enumerate}
In the descriptive configuration notation, we have here added lower-case superscripts to indicate where a defect is located: '$t$' stands for the 'top' of the shell. The top is also the thinnest point, as the aqueous F127 solution is less dense than the LC. 

In total we have studied hundreds of shells similar to those in Fig.~\ref{fgr:5cbshells}. In freshly produced shells at room temperature we always find all three configurations, with a strong preference for T1 [4(+\sfrac{1}{2}$^t$)]. In an experiment with a total of 55 pristine shells we counted 40 in T1, 9 in T2 and 6 in T3 configuration.

As we heat and then cool the shells, we identify several surprising features in the textural evolution, with some differences depending on shell type. We briefly go through the key observations here, in the order of the starting configurations, T1--T3. This means that we will discuss first the right shell (blue arrows), then the left shell (red arrows), and finally the middle shell (green arrows) in Fig.~\ref{fgr:5cbshells}. In Section~\ref{E7-F127-shells}, we go through each trajectory in more detail, based on the expanded temperature range observations in E7 shells.

\begin{enumerate}

    \item In all cases, a shell changes gradually on heating from the original T configuration, via H, and finally to N (Fig.~\ref{fgr:5cbshells}a-n). During the T${}^{\phantom{1}\rightarrow}_{\textrm{heat}}$H transition we can distinguish two or three characteristic steps, depending on the starting configuration. 
    \begin{enumerate}
    
    \item Starting with the T1 shell [4(+\sfrac{1}{2}$^t$), blue arrow], two of the four +\sfrac{1}{2} defects first move to the top (Fig.~\ref{fgr:5cbshells}b--c) and remain there. They are very close but do not yet fuse. We define this as \textbf{Step 1} of the T1-initiated trajectory. The other two defects first move apart towards the equator (Fig.~\ref{fgr:5cbshells}b--d), then approach each other on the bottom side of the shell (e). Between Fig.~\ref{fgr:5cbshells}f--g the two top +\sfrac{1}{2} defects fuse into a single +1 defect, with the sudden appearance of a spiral pattern in POM signifying a twist in the director field in the top half of the shell. We define this as \textbf{Step 2}, occurring before the remaining +\sfrac{1}{2} defects have met at the bottom. They do so in \textbf{Step 3} (just before Fig.~\ref{fgr:5cbshells}i, see Movie 1), immediately fusing and leaving a second +1 point defect antipodal to the first +1 defect. Again, the defect fusion is connected to the sudden appearance of a spiral pattern, this time on the bottom shell half, hence the director field twist now extends throughout the shell. The end configuration is thus a hybrid +1$^b$,+1$^t$ (superscript '$b$' for 'bottom'). 
    
    The three steps of this trajectory are illustrated graphically from multiple perspectives in Fig.~\ref{fgr:3steps}. It is here easy to see that Step 1 induces no twist (panels a--c), whereas between Step 2 and Step 3 (d--f) a twist is localized to the top half of the shell. After Step 3, finally (g--i), the director field twists throughout the shell. While the drawing is done for one twisting sense, the sense in experiments varies randomly between shells, as expected without molecular chirality.

\begin{figure}
\centering
 \includegraphics[width=8.5cm]{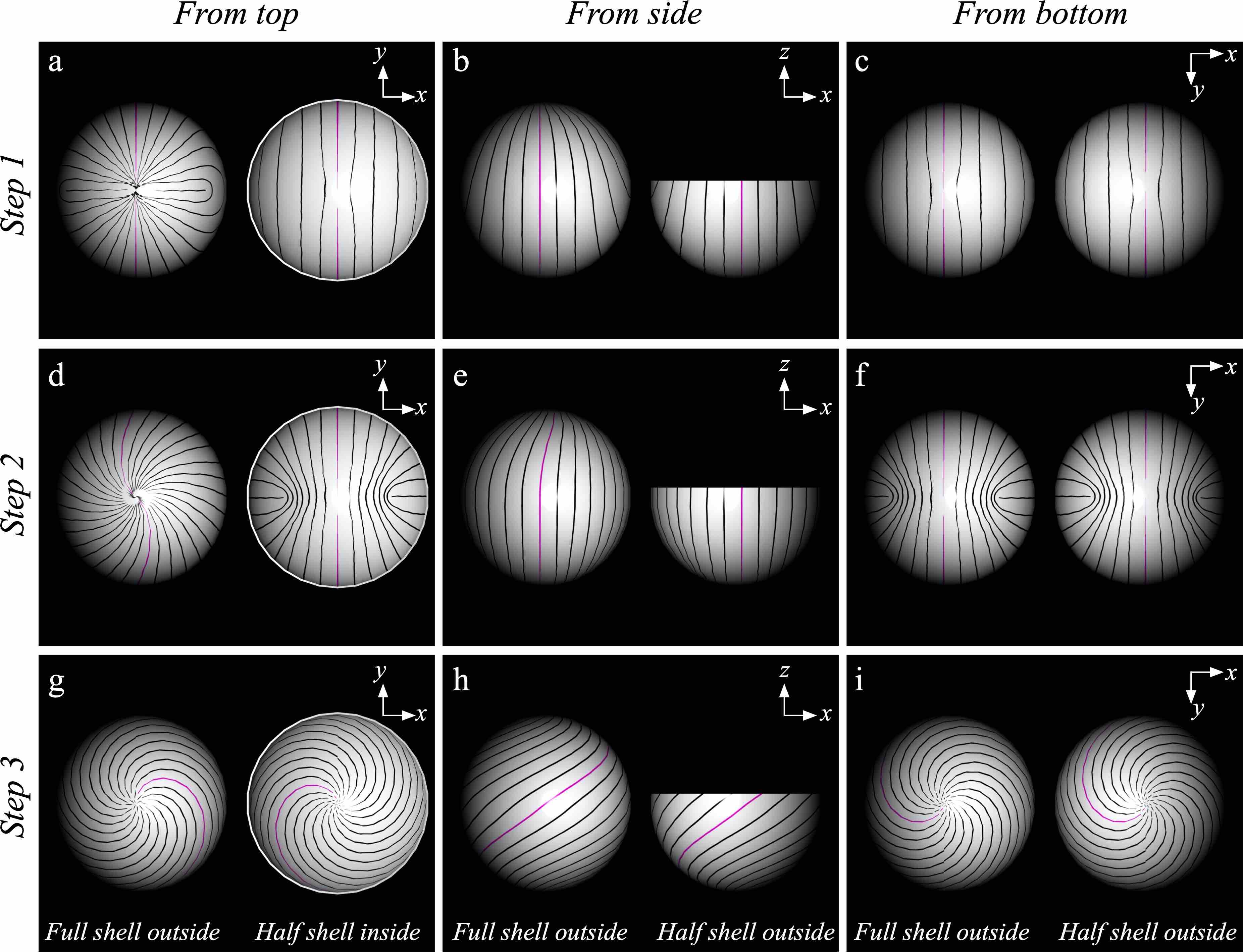}
 \caption{Computer-drawn illustrations of how the tangential director field projection changes in shells following the 3-step realignment trajectory that begins with the T1 [ 4(+\sfrac{1}{2}$^t$)] configuration. In each panel, one full shell (left) and one half shell (right) is drawn, both with identical tangential director fields mapped onto them. One field line has been highlighted in pink color, to facilitate tracing the director around the shell surface. The shells are viewed along three perspectives, from the top in the left column, thus looking into the half shell, from the side in the middle column, and from the bottom in the right column, viewing both shells from their outsides. The drawings correspond to the (predominantly) tangential shell side immediately following Step 1 (first row), Step 2 (middle row) and Step 3 (bottom row), respectively. The perspective in the left-most column corresponds to that in upright microscopy, the full shell showing the texture obtained with focus on the top and the half shell the texture when the focus is at the bottom. In (a), the two +\sfrac{1}{2} defects at the shell top, gathered very close without merging into a +1 defect, cannot be well resolved with the graphics software used.}
 \label{fgr:3steps}
\end{figure}
        
    \item Next, we study the closely related trajectory starting from T2 [2(+\sfrac{1}{2}$^t$),+1$^t$]. This trajectory exhibits the same final steps as defined above, but Step 1 is absent, since only two +\sfrac{1}{2} defects are present from the beginning. As we heat from room temperature, the initial +1 defect moves up to the top of the shell, where it stops (Fig.~\ref{fgr:5cbshells}c). Soon afterwards we see the characteristic spiral pattern revealing a director twist in the top shell half, see Fig.~\ref{fgr:5cbshells}d, constituting \textbf{Step 2}. As above, the two +\sfrac{1}{2} defects first move apart towards the equator, which they reach at about the temperature of Step 2. On continued heating, they then approach each other on the bottom side of the shell, where they eventually fuse into a +1 point defect (\textbf{Step 3}) that is antipodal to the original one, see Fig.~\ref{fgr:5cbshells}f. Again, Step 3 coincides with the sudden appearance of a spiral pattern on the bottom half of the shell, showing that the director field twist now extends throughout the shell. The stable hybrid configuration thus again ends up +1$^b$,+1$^t$.

    \item The T3 shell [2(+1$^t$)], finally, gives rise to a realignment sequence that is altogether different from the two trajectories described above. We identify three key steps here, allowing us to compare the temperatures at which these steps occur in 5CB and E7 shells, but we note that they are not necessarily directly comparable to the steps seen in the trajectories starting from T1 or T2. In the study of the T$\rightarrow$H transition upon stabilizer exchange by Lopez-Leon and Fernandez-Nieves \cite{lopez2009topological}, the starting configuration was also 2(+1$^t$), and the trajectory we observe in the T3 shell matches one of the trajectories described in their report. A defect ring first arises along a great circle between the two +1 point defects, which soon start leaving the top, moving towards the equator along the ring in opposite directions (Fig.~\ref{fgr:5cbshells}c--d). In this trajectory, we call the start of the defect movement \textbf{Step 1}. Somewhat later, the ring leaves the great circle location, moving towards one side and shrinking in the process (Fig.~\ref{fgr:5cbshells}i). Initially both point defects move with the ring, but, as seen in Movie 1, soon one defect detaches (we denote this \textbf{Step 2} of this trajectory), moving up to the top of the shell. In the experiment of Fig.~\ref{fgr:5cbshells} and Movie 1 the transition to fully normal alignment (see below) occurs in the T3-initiated shell prior to completion of the T${}^{\phantom{1}\rightarrow}_{\textrm{heat}}$H transition by \textbf{Step 3}, hence we will come back to the final stage in Section~\ref{e7_2(+1)}. The extended temperature range of E7 shells there allows us to follow the T${}^{\phantom{1}\rightarrow}_{\textrm{heat}}$H transition until the end.

    \end{enumerate}

%disclinations, connecting the shell in- and outsides, are not allowed in hybrid alignment since one side is defect-free

    \item On yet further heating, the shell turns from hybrid to fully normal (H${}^{\phantom{1}\rightarrow}_{\textrm{heat}}$N), see Fig.~\ref{fgr:5cbshells}j--k. This happens less than 1~K below $T_{NI}$. Continuing to heat to $T_{NI}$, the shell turns isotropic (Fig.~\ref{fgr:5cbshells}l).
    
    \item Cooling the shells back to nematic, the alignment transitions take place in reverse: N${}^{\phantom{1}\rightarrow}_{\textrm{cool}}$H with twisted director field (Fig.~\ref{fgr:5cbshells}$\alpha-\gamma$), finally T after a few minutes at room temperature (Fig.~\ref{fgr:5cbshells}$\delta-\zeta$).
    
    \item{Significantly, the final T state (Fig.~\ref{fgr:5cbshells}$\zeta$) shows only \textit{one} type of tangential director field: T3 [2(+1$^t$)]. No single shell displays any +\sfrac{1}{2} disclination after the slow cooling process of Fig.~\ref{fgr:5cbshells}$\alpha$--$\zeta$, thus T1 and T2 configurations are entirely absent. In the quantitative experiment mentioned above, all 55 shells remained intact after the slow cooling and all were in T3 configuration. We will propose an explanation for this observation, which we believe is intimately linked to the intermediate H configuration upon slow cooling, in Section~\ref{finalconfigurationexplanation}.}
    
\end{enumerate}

\subsubsection{Simulation of stable hybrid configurations}
\begin{figure}
\centering
 \includegraphics[width=8cm]{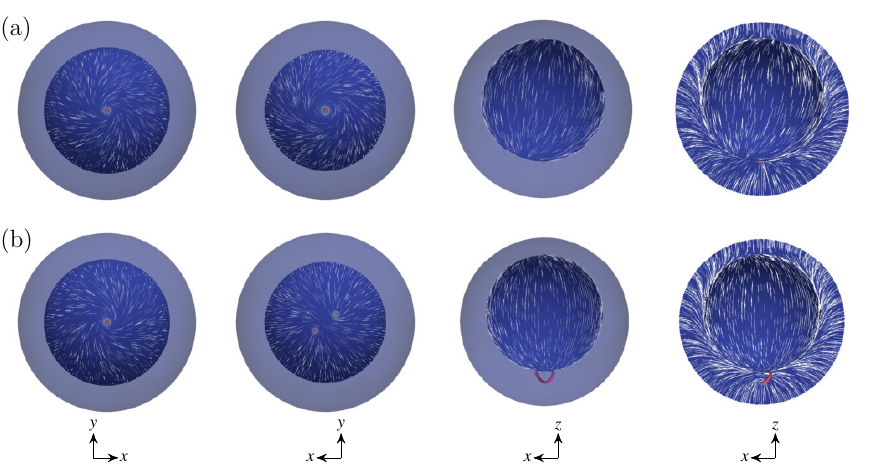}
 \caption{Two (meta-) stable H states (tangential inside, normal outside) obtained from numerical simulations, with the reduced temperature (different from the experimentally defined $T_r$, used below) $t = -1.79, and c= 0.1, \rho = 0.7, \eta = 4$. The shells are viewed from the thinnest part (shell top), thickest part (bottom) and side, respectively, from left to right. The final two columns have the same side view perspective, the third column showing the director field only at the inner shell boundary while the fourth column shows the director field within the shell. With two antipodal +1 defects (a; +1$^b$,+1$^t$ configuration) the director field is twisted from top to bottom, whereas a shell with a 2(+\sfrac{1}{2}$^b$),+1$^t$ configuration (b) exhibits twist only on the top half, where the +1 defect resides. Both +\sfrac{1}{2} defects at the bottom are on the inside, connected by a U-turned disclination line. Calculations of the LdG free energy show that +1$^b$,+1$^t$ is the global minimum whereas 2(+\sfrac{1}{2}$^b$),+1$^t$ is a local minimum.} 
 \label{fgr:hybrid}
\end{figure}

 We first focus on the stable H configuration, see Fig.~\ref{fgr:hybrid}, setting the inside boundary condition tangential and the outside normal; this is compatible with experimental data above and identical to the starting configuration in Section~\ref{hybriddetails}. To compare our numerical results with earlier studies, and to provide initial T configurations for simulated heating experiments, we also simulate tangential director fields in thin and thick shells; see Appendix~\ref{tangentialsim}.

Regarding the final step in observation 1 in the above list, we find multiple locally stable configurations but the experimentally detected +1$^b$,+1$^t$ H configuration has the lowest free energy for reasonable parameter values $1/\xi_R = 50$ ($R \approx 1 \mu \mathrm{m}$) and $\eta = 4$ [$t = -1.79, c= 0.1, \rho = 0.7$]. This conclusion holds generally for large $1/\xi_R$ and large $\eta$. The two +1 point defects on the inner, tangential interface move to antipodal points (as far away from each other as possible within geometrical constraints), to minimize the distortion energy of the director field.%The 2(+\sfrac{1}{2}$^b$),+1$^t$ configuration might be unstable when the shell diameter is as large as that in the experiments.

Particularly interesting is that we find a twist throughout the shell for the +1$^b$,+1$^t$ H configuration (Fig.~\ref{fgr:hybrid}a), whereas a 2(+\sfrac{1}{2}$^b$),+1$^t$ H configuration (Fig.~\ref{fgr:hybrid}b) has the twist localized to the top half, hosting the +1 defect, in good agreement with experimental observations (visualized in Fig.~\ref{fgr:3steps}d--f and g--i, respectively). Moreover, the simulations demonstrate that an H shell can exhibit +\sfrac{1}{2} disclinations, albeit of a different geometry than in T shells. The disclinations running from the shell inside to the outside in the latter require both interfaces to be tangential; if the director were tilted out of the interface plane near the disclination, the 180$^\circ$ rotation around a +\sfrac{1}{2} defect would invert the tilting direction, thus it would no longer be a symmetry operation. However, +\sfrac{1}{2} disclinations can still exist as long as \textit{one} interface is tangentially aligned, by curving the disclination line into the topology of a U, see Fig.~\ref{fgr:hybrid}b, right. Visible +\sfrac{1}{2} defects on H shells are then connected in pairs, comprising the start and end points of a U-turned disclination on the tangential boundary (the inner boundary in this case). %Both are located on the tangential side, here the inside. This situation must prevail during the experimentally observed T$\rightarrow$H transition when +\sfrac{1}{2} defects are still to be seen.

\subsubsection{Simulation of the H$\rightarrow$T trajectory}
\begin{figure}
\centering
 \includegraphics[width=8cm]{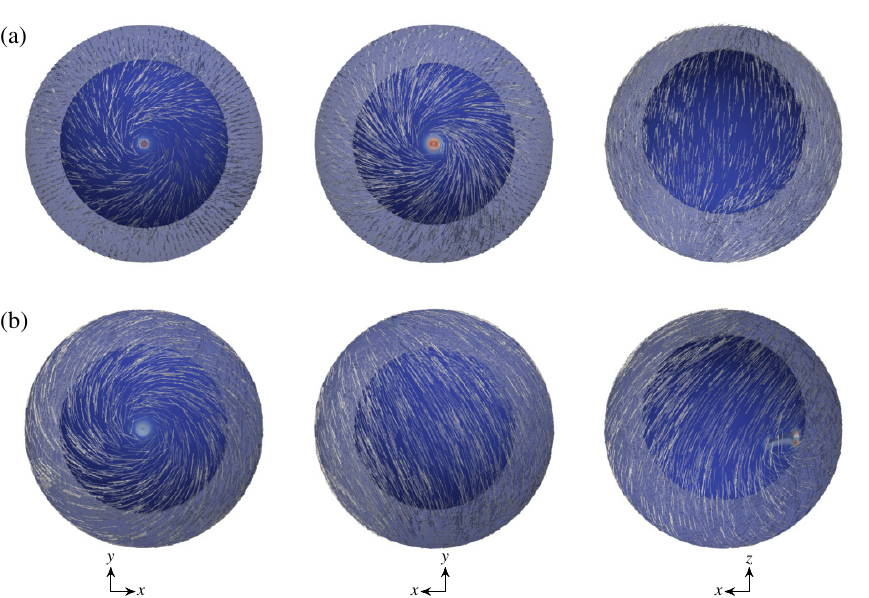}
 \caption{Simulated trajectory of the H$\rightarrow$T transition starting from a twisted H configuration [$t = -1.79$, $c = 0.1$, $\rho = 0.7$ and $\eta = 4$]: (a) n = 1000, (b) n = 30000, where $n$ denotes the number of iterations. Left to right: view from the thin part, the thick part and side, respectively.} 
 \label{fgr:H_P}
\end{figure}

Regarding observation 4, we perform a gradient flow simulation at fixed temperature for a shell that has been set up with the equilibrium +1$^b$,+1$^t$ H configuration, switching to tangential anchoring at both interfaces at the start of the simulation. The results for $c = 0.1$, $\rho = 0.7$ and $1/\xi_R = 50$ ($R \approx 1 \mu \mathrm{m}$) with $\eta = 4$ at $t = -1.79$ are shown in Fig. \ref{fgr:H_P}. Initially, the $+1$ defect near the thinnest part of the shell appears stable, now as a $+1$ defect pair (one defect on the shell inside, one on the outside), see Fig. \ref{fgr:H_P}a. At a later point, the + 1 defect in the thick part breaks into two +\sfrac{1}{2} defects, which move towards the thinnest part of the shell, see Fig. \ref{fgr:H_P}b. Eventually, the remaining +1 defect pair becomes unstable, yielding a final $4(+\sfrac{1}{2}^t$), or T1, state.

The disagreement with the experimental results in Fig.~\ref{fgr:5cbshells} concerning the final stable configuration may have several explanations. While it could be related to the limited size of the simulated shell, a more important factor is probably that the experiment follows a trajectory through which the order parameter and, consequently, elastic constants, interfacial tension and boundary conditions vary continuously (see Discussion). As will become clear below, it is very likely that the alignment at one or both interfaces changes gradually during the experiment, thus passing through a region where a tilted alignment (neither tangential nor normal) is promoted, with possible exceptions in localized regions. In contrast, the simulation is done for constant temperature deep in the nematic phase, fixed elastic constants, and a suddenly imposed alignment change, from perfect H to perfect T conditions. Thus, the simulations find minima other than those observed in the experiments. Nevertheless, the numerical experiment captures the transient stability of at least one $+1$ defect pair and the tendency of all topological defects to migrate towards the thinnest part of the tangential shell. As will become clear in the Discussion, this simulation may be more relevant for the conditions that led to the appearance of the pristine T1 shells.

\subsection{Detailed analysis of tangential--hybrid--normal trajectories in E7 shells}\label{E7-F127-shells}

In the realignment trajectories described in Fig.~\ref{fgr:5cbshells}, Steps 2 and 3 always occur above the critical micelle temperature (24$^\circ$C) of a 1\% aqueous F127 solution \cite{Alexandridis1994}, hence we do not expect changes in the behavior of F127 to be at the origin of these steps. Moreover, the temperatures of 5CB shell realignment transitions discussed above are largely below the range in which realignment was detected in our preliminary study on 8CB shells; the relevant range in that case was 33--40$^\circ$C \cite{liang2011liquid}. These observations suggest that the realignment is not primarily due to thermally induced effects in the F127 solution but rather related to the variation of nematic order within the LC. To confirm this, we conduct experiments on shells prepared using E7 (supplementary Movies M2--M4, Figures \ref{fgr:fourdefects}, \ref{fgr:threedefects} and \ref{fgr:twodefects}), with a much higher $T_{NI}$ than 5CB.

An added benefit of the E7 shells is that the expanded nematic temperature range allows us to study each trajectory in more detail and, in contrast to the 5CB shell case, there are no overlapping transitions, not even when starting in a T3 configuration. As above, all shells are pristine at the beginning of the experiment, i.e., they have not been subject to temperature changes after the rapid cooling ending production. We only illustrate one shell at a time but similar experiments have been carried out on 20--50 E7 shells for each starting configuration, allowing us to be reasonably confident about the generality of our observations and conclusions. For selected panels we provide corresponding director field sketches. We refer to the right part of Fig.~\ref{fgr:5cbshells} for interpretation of the textures in general. 

\subsubsection{Realignment from T1 configuration [4(+\sfrac{1}{2}$^t$)]}\label{start4discs}
\begin{figure}
\centering
 \includegraphics[width=8cm]{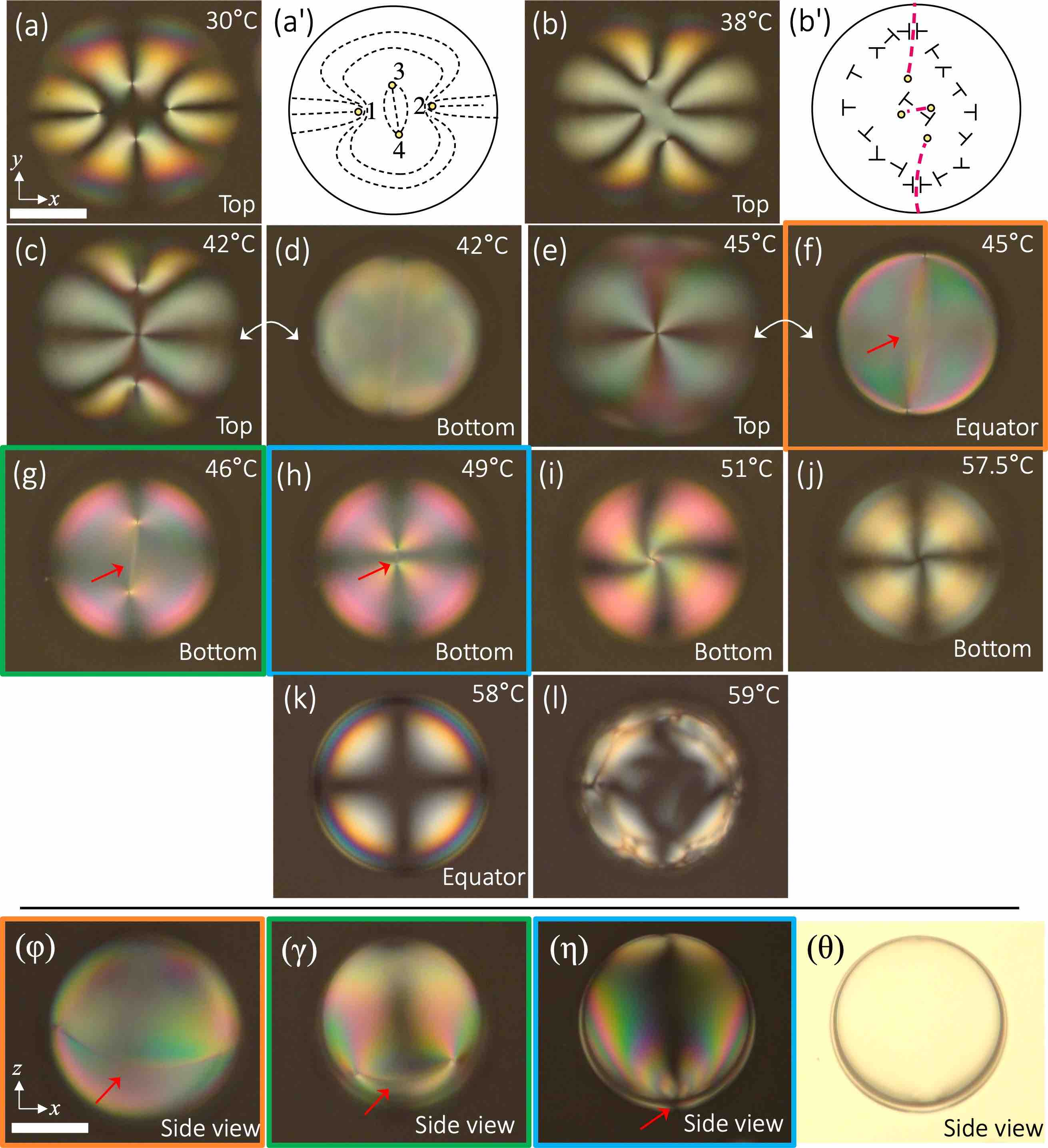}
 \caption{Alignment transformation upon heating of an E7 shell starting in T1 configuration [4(+\sfrac{1}{2}$^t$)], stabilized by F127 on both sides. The shell is between crossed polarizers (horizontal and vertical) except in panel ($\theta$), obtained without analyzer. The focal plane is noted in each top view panel (a--l, coordinate system in a). Panels ($\phi$), ($\gamma$), ($\eta$) and ($\theta$) show side views (coordinate system in $\phi$), obtained with the microscope tilted 90$^\circ$; ($\phi$), ($\gamma$) and ($\eta$)  roughly correspond to (f), (g) and (h), respectively, as illustrated by the colored frames (for practical reasons, top and side views are from separate experiments). The ground state director field is drawn in (a$'$). In (b$'$), the tilt direction is suggested with nails, nail head and tail signifying upwards- and downwards-pointing director, respectively (or vice versa). The $\pi$ defect lines are drawn as dashed red lines in b$'$ and highlighted with red arrows in the micrographs. Scale bar: 50 $\mu$m.}
 \label{fgr:fourdefects}
\end{figure}

The four initial defects are labelled 1--4 in Fig.~\ref{fgr:fourdefects}a$'$, corresponding to the micrograph in (a). To map out the director field with certainty, the shell is observed temporarily with a first-order $\lambda$-plate (see Movie M2). As \textbf{n} on the shell outside starts tilting upon heating (Fig.~\ref{fgr:fourdefects}b--d), two faint lines appear. A short line (red dashed line in Fig.~\ref{fgr:fourdefects}b$'$) connects defects 1--2 over the shell top, and a longer (red dashed line in (b$'$), highlighted by red arrows in micrographs) connects defects 3--4, running around the back of the shell. These lines can be interpreted as $\pi$-lines (see also Section~\ref{whyalignmentchange}), separating regions with opposite directions of interface tilt, and thus of bend within the shell, during the  T${}^{\phantom{1}\rightarrow}_{\textrm{heat}}$H transition. We indicate the tilt with nails in Fig.~\ref{fgr:fourdefects}b, where the nail head and tail can represent upwards- and downwards-pointing \textbf{n}, respectively. However, as we cannot experimentally distinguish between the two possible tilt directions, the inverse interpretation is equally possible.  

As heating continues, defects 1 and 2 move closer until we reach Step 1 (Fig.~\ref{fgr:3steps}a--c), where the defects settle at the shell top without merging (Fig.~\ref{fgr:fourdefects}e). They are so close to each other that their separation can only be distinguished with perfect focus (Movie M2). Defects 3 and 4 initially move further apart, accelerating after Step 1, reaching the equator in (f) and the bottom region in (g). They merge (Step 3, Fig.~\ref{fgr:3steps}g--i)) between (h) and (i), immediately triggering a twisting of the surrounding bottom shell half director field recognized by a spiral pattern. Defects 1 and 2 must have merged into a +1 defect slightly earlier (Step 2, Fig.~\ref{fgr:3steps}d--f)), although the event is not visible in Fig.~\ref{fgr:fourdefects}/Movie M2 because the focus stays at the shell bottom from (g) to (i). (It is seen, however, in two other E7 shells that start from the T1 configuration in Movie M4.) In Fig.~\ref{fgr:fourdefects}i, the shell thus has a hybrid +1$^b$,+1$^t$ configuration with twisted director field, the tangential side having the director field illustrated in Fig.~\ref{fgr:3steps}g--i. The antipodal +1 defects disappear around 58$^\circ$C (Fig.~\ref{fgr:fourdefects}j--k) when the shell becomes fully normally aligned (H${}^{\phantom{1}\rightarrow}_{\textrm{heat}}$N). The shell clears at about 59$^\circ$C (Fig.~\ref{fgr:fourdefects}l).

Panels ($\phi$--$\theta$) in Fig.~\ref{fgr:fourdefects} show representative micrographs of another shell initially in T1 configuration undergoing the same realignment trajectory upon heating, viewed from the side by tilting the microscope 90$^\circ$. It is experimentally impossible to have both views at the same time, hence the shell in the side view in ($\phi$), corresponding to the top view (f), has the defect line nearer the equator than the shell in (f). The defect locations in the side views in ($\gamma$--$\eta$) correspond quite well to those in the top views in (g--h), respectively. Panel ($\theta$) shows the side view shell without analyzer, making it easy to see the asymmetry in shell thickness.

\begin{figure}
\centering
 \includegraphics[width=8cm]{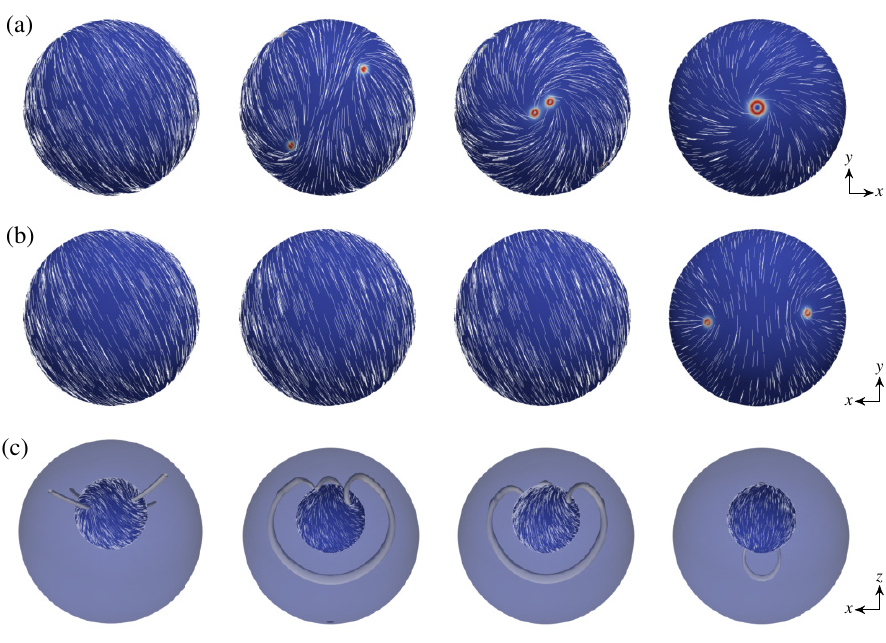}
 \caption{Simulated trajectory of the T$\rightarrow$H transition starting from T1 [4(+\sfrac{1}{2}$^t$)], with $t = -1.79$, $c = 0.2$, $\rho = 0.4$ and $\eta = 4$. We plot \textbf{n} at the inner shell boundary [left to right: simulations steps $n$ = 0, 10000, 15000 and 20000]: (a) view from thin part, (b) view from thick part, (c) side view. Defects are identified with the red regions or regions of high biaxiality.}\label{fgr:4_P_H}
\end{figure}

%\subsubsection{Starting from 4(+\sfrac{1}{2}$^t$); results from simulations}\label{Simul4(+1/2)}
In Figure~\ref{fgr:4_P_H}, we employ the gradient flow algorithm to simulate the transient dynamics within a simplified framework, tracing a possible path towards a stable H state. The simulation is done for fixed temperature and material parameters and we do not consider tilted alignment at any interface, hence the outside boundary condition is suddenly switched to normal while the inside boundary remains tangential. The shell is asymmetric with thin top, outer radius of $1$~$\mu$m and LC elastic anisotropy parameter, $\eta=4$. 

The transition starts with two +\sfrac{1}{2} defects on the shell inside connecting via a short U-curved disclination line through the thin top of the shell, and the two others via a long curved disclination line extending around the back, see Fig.~\ref{fgr:fourdefects}c. Similar to the experiment, the first two approach each other, moving towards the top. However, upon meeting they merge immediately to yield a +1$^t$ defect, hence in the simulation Step 1 and Step 2 coincide. The twisting of the simulated director field starts when the defects are very close, but before the merger, in contrast to experiment, see field plots 2 and 3 in Fig.~\ref{fgr:4_P_H}a. %The slight differences between experiments and simulation are likely due to the fact that the simulation does not capture the gradual changes in parameters taking place in the experiment. Most conspicuously, the simulation has perfectly normal outer boundary conditions, whereas the real situation most likely includes tilted alignment, approaching normal anchoring on the outside. 
The remaining two +\sfrac{1}{2} defects migrate to the thickest part and settle there, sufficiently separated so as not to induce a twisted director field in their shell half. In other words, we do not reproduce Step 3 of the experiment nor do we reach the final +1$^b$,+1$^t$ hybrid configuration with a twisted director field throughout the shell. 

We identify three main reasons for the differences between experiments and simulations regarding the trajectory as well as the final state. Firstly, the simulations are not carried out as a function of temperature, instead triggering the realignment by switching the outer boundary from hard tangential to hard normal, while the inside remains hard tangential throughout. Secondly, compared to the experimental shells, the simulated shells are small and relatively thick, since computational resolution renders it difficult to extract structural information for thinner shells. Finally, these complex systems have multiple stable configurations \cite{Yin2020prl} and the gradient flow algorithm may converge to a locally stable configuration. Indeed, our direct simulations of hybrid director fields in Fig.~\ref{fgr:hybrid} produce a +1$^t$,2(+\sfrac{1}{2}$^b$) configuration with twisted director field through the shell top and a +1$^b$,+1$^t$ state with twist throughout the shell, the latter corresponding to the global free energy minimum. As the +1$^t$,2(+\sfrac{1}{2}$^b$) state is a local energy minimum, our dynamical algorithm converges to it and stays there, unable to overcome the energy barrier needed to reach the +1$^b$,+1$^t$ state.

\subsubsection{Realignment from T2 configuration [2(+\sfrac{1}{2}$^t$),+1$^t$]}\label{start2discs+1point}
 
\begin{figure}
\centering
 \includegraphics[width=8cm]{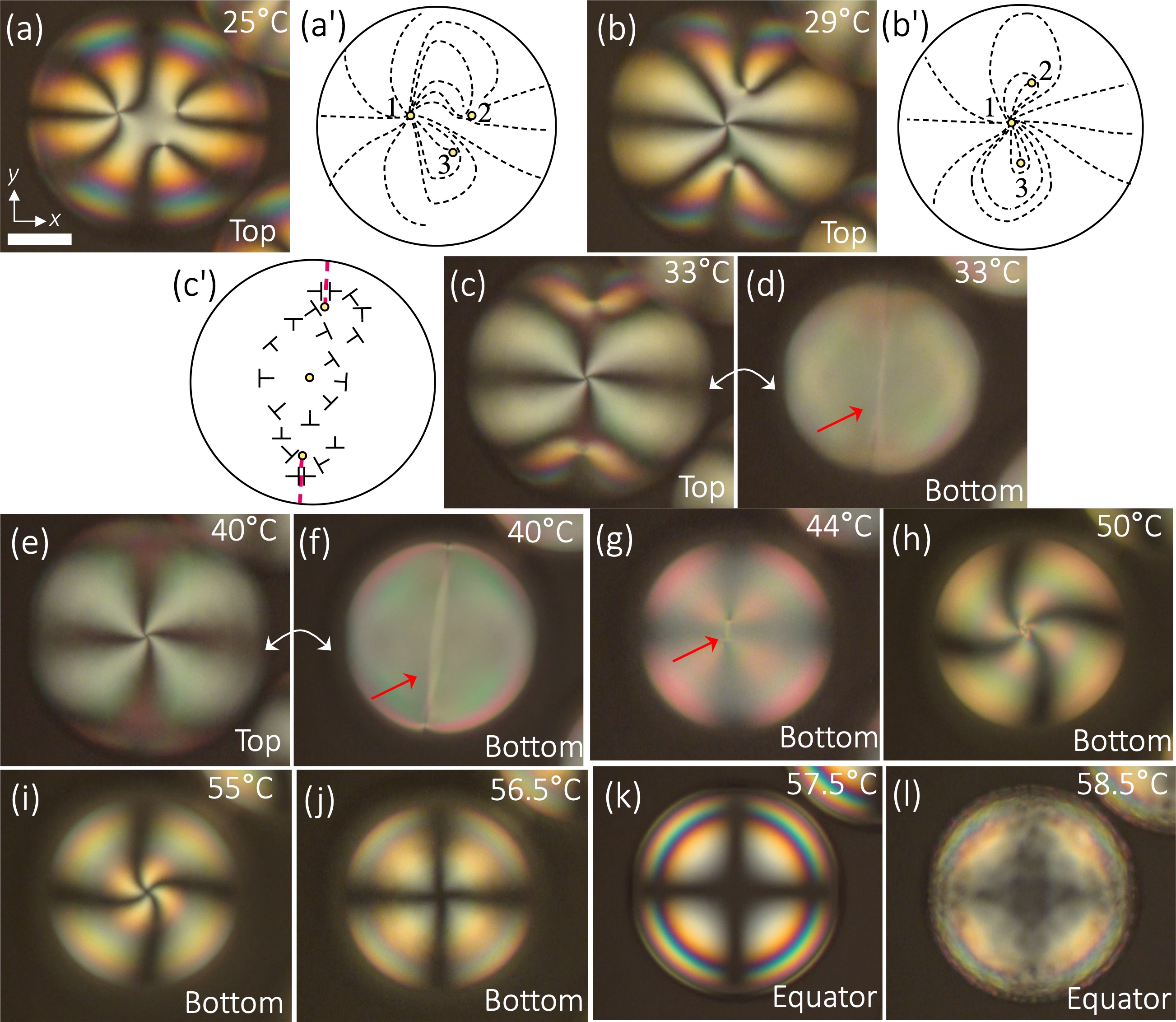}
 \caption{Alignment transformation upon heating (a--l) of an E7 shell starting in the T2 configuration [2(+\sfrac{1}{2}$^t$),+1$^t$], stabilized by F127 on both sides. The shell is between crossed polarizers (horizontal and vertical). The focal plane is noted in each panel. We draw the ground state director field (a$'$) and the new one (b$'$) arising after a defect exchange that initiates the transformation. In (c$'$), the tilt direction is suggested with nails, nail head and tail signifying upwards- and downwards-pointing director, respectively (or vice versa). The $\pi$ defect line is highlighted with red arrows in the micrographs and its ends reaching the top shell half are drawn as dashed red lines in c$'$. Scale bar: 50~$\mu$m.}
 \label{fgr:threedefects}
\end{figure}

 In this realignment sequence, a preliminary step  is a brief merger and immediate re-splitting of the +1 and one of the +\sfrac{1}{2} defects (1 and 2 in Fig.~\ref{fgr:threedefects}a$'$), rearranging the surrounding director field as sketched in Fig.~\ref{fgr:threedefects}b$'$ ($\lambda$-plate appearance in Movie M3). The resulting +1 defect moves towards the top, reaching it in panel (c/c$'$) with the director field twist in the top half of the shell immediately seen (Step 2, Fig.~\ref{fgr:3steps}d--f; no Step 1 in this sequence). The remaining trajectory, Fig.~\ref{fgr:threedefects}d--l, is very similar to that of Section~\ref{start4discs}.

\begin{figure}
\centering
 \includegraphics[width=8cm]{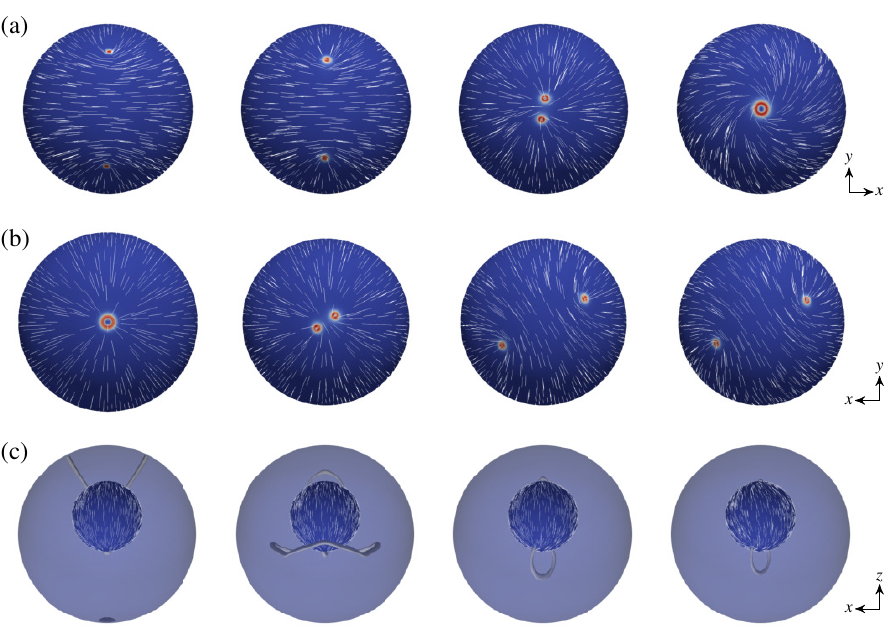}
 \caption{Simulated trajectory of the T$\rightarrow$H transition starting from T2 [2(+\sfrac{1}{2}$^t$),+1$^t$], with $t = -1.79$, $c = 0.2$, $\rho = 0.4$ and $\eta = 4$. We plot \textbf{n} at the inner shell boundary [left to right: simulation steps $n$ = 0, 5000, 20000 and 40000]: (a) view from thin part, (b) view from thick part, and (c) side view. }\label{fgr:3_P_H}
\end{figure} 

%\subsubsection{Starting from 2(+\sfrac{1}{2}$^t$),+1$^t$; results from simulation}\label{Simul+1,2(+1/2)}
The corresponding gradient flow dynamics simulation (asymmetric shell, dimensions and simulation conditions as above) produces quite a different trajectory (Fig.~\ref{fgr:3_P_H}) compared to the experiment, in particular for the original +1 point defect on the inner surface. It first moves to the thick shell side where it splits into two +\sfrac{1}{2} defects connected by a wide U-turned disclination loop, see Fig. \ref{fgr:3_P_H}b. The two original +\sfrac{1}{2} defects (connected by a U-turned disclination loop after the outer interface alignment is switched to normal) move to the thinnest point, where they merge into a +1 defect with a twisted director field around it, see row (a). Their behavior is thus closer to what is seen in experiment, albeit on the opposite shell side. 

The final H state is again +1$^t$,2(+\sfrac{1}{2}$^b$), since the two +\sfrac{1}{2} defects on the thick side of the shell do not merge. As above, the discrepancies compared to the experimental observations are most likely due to the simplifications of the simulation conditions and to the gradient flow simulation getting stuck in a local energy minimum. %Whilst we expect the $+1^t, +1^b$ hybrid state to be energy minimizing for thin hybrid shells, as discussed in the context of Figure~\ref{fgr:hybrid}, the numerical gradient flow algorithm can converge to the competing $+1^t, 2(+1/2^b)$ configuration, constituting a local energy minimum. It can stay there because the free energy difference between the $+1^t, +1^b$ and $+1^t, 2(+1/2^b)$ configurations decreases for thicker shells as used in the simulation.
%this is expected because transient dynamics are not unique. 
%There are typically multiple transition pathways for complex solution landscapes. Numerically, we observe the two +\sfrac{1}{2} defects merging to yield a twisted +1 defect in the thinnest part of the tangential shell, whilst the two +\sfrac{1}{2} defects migrate and localise near the thickest part. The final state is the $1$-twisted state in Section IIIB.
%Simulated trajectory of tangential--hybrid transition starting from one +1 and two +\sfrac{1}{2} defects is shown in Fig. \ref{fgr:3_P_H}.
%In this case, two +\sfrac{1}{2} defects move to the thinnest part of the shell and merge into a twist +1 defects, while the “+1 defects” moves to the thickest part first and split into two +\sfrac{1}{2} defects.

\subsubsection{Realignment from T3 configuration [2(+1$^t$)]}\label{e7_2(+1)}
\begin{figure}
\centering
 \includegraphics[width=8cm]{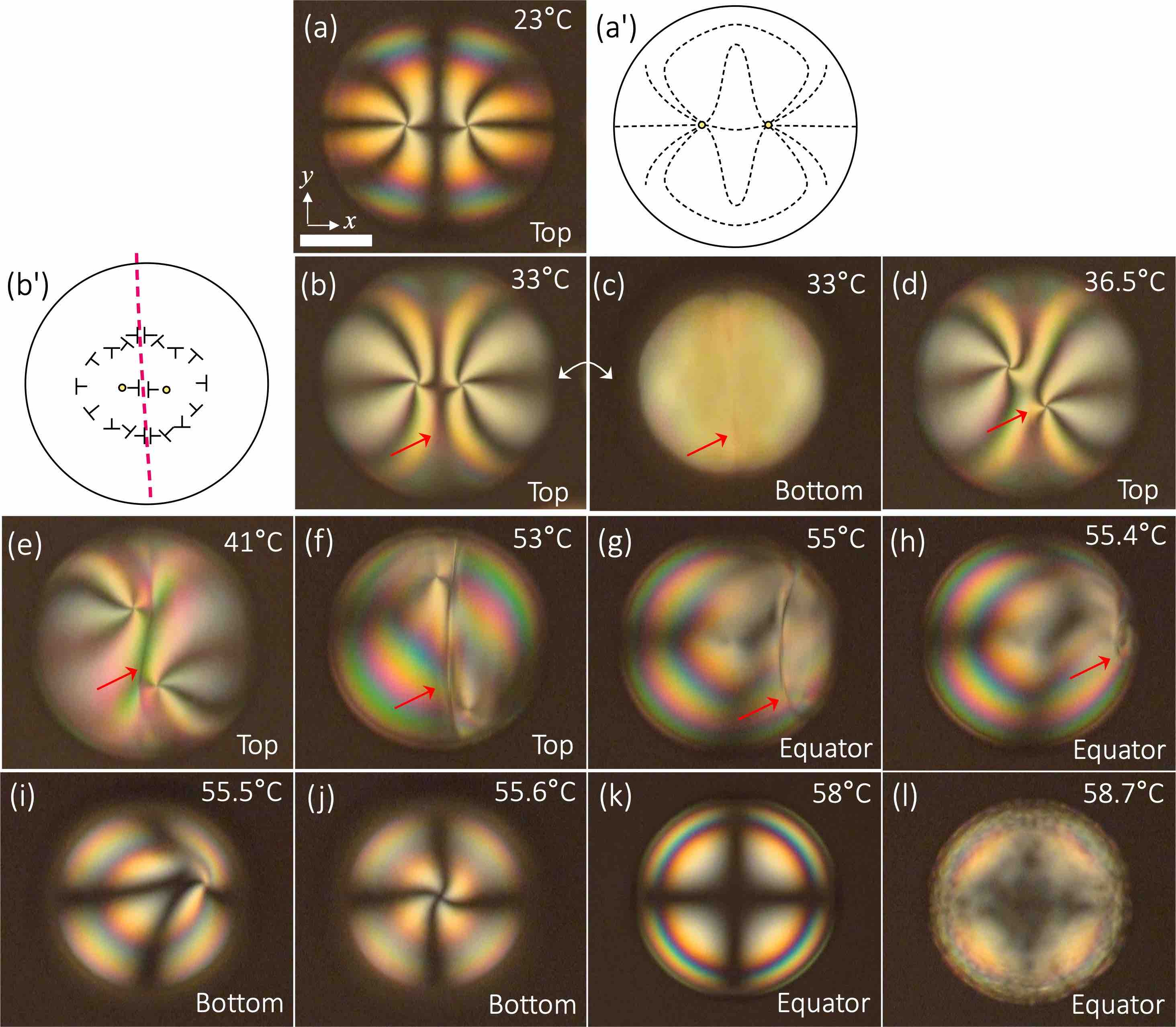}
 \caption{Alignment transformation upon heating (a--l) of an E7 shell starting in T3 configuration [2(+1$^t$)], stabilized by F127 on both sides. The shell is between crossed polarizers (horizontal and vertical). The focal plane is noted in each panel. The ground state director field is drawn in (a$'$). In (b$'$), the tilt direction is suggested with nails, nail head and tail signifying upwards- and downwards-pointing director, respectively (or vice versa). The $\pi$-circle separating opposite tilt directions is drawn in red in b$'$ and highlighted with red arrows in the micrographs. Scale bar: 50~$\mu$m.}
 \label{fgr:twodefects}
\end{figure}

Compared to the cases where +\sfrac{1}{2} disclinations are initially present, the first signs of response of an E7 shell initially in T3 configuration (Fig.~\ref{fgr:twodefects}a/a$'$) are seen at a later stage, just like for the corresponding 5CB shells (Fig.~\ref{fgr:5cbshells}). By about 33$^\circ$C the colors have changed and a $\pi$ defect circle can be recognized, see Fig.~\ref{fgr:twodefects}b/b$'$--c and Movie M4. The circle arises half-way between the two +1 defects, aligned normal to the line separating them. At 35$^\circ$C (between panels (c) and (d) in Fig.~\ref{fgr:twodefects}) the two +1 defects start moving in opposite directions along the $\pi$-circle. As for the case of 5CB shells, we define this event as Step 1 in the T3-initiated trajectory. We remind that the three steps starting from T3 are not directly comparable with those starting from T1 or T2. The defects move further apart (Fig.~\ref{fgr:twodefects}e) and at 53$^\circ$C (Fig.~\ref{fgr:twodefects}f) the circle starts moving rightwards, pushing one of the +1 defects ahead of it. The other +1 defect detaches (Step 2) and retracts to the shell top in panel (g). 

At 55.4$^\circ$C (Fig.~\ref{fgr:twodefects}h) the $\pi$-circle has almost closed in upon itself, the trapped +1 defect pushed to the far right. Soon after, the circle disappears (Step 3), allowing the previously trapped defect to move down to the bottom of the shell (Fig.~\ref{fgr:twodefects}i--j). In contrast to the trajectory with 5CB shells, we here obtain the usual +1$^b$, +1$^t$ H configuration at 55.6$^\circ$C (Fig.~\ref{fgr:twodefects}j), with a twisted director field throughout the shell. Heating yet a bit more, the shell acquires an N configuration at 58$^\circ$C (Fig.~\ref{fgr:twodefects}k). The shell clears at 58.7$^\circ$C (Fig.~\ref{fgr:twodefects}l). For all shells starting in T3, whether E7 or 5CB, our T${}^{\phantom{1}\rightarrow}_{\textrm{heat}}$H trajectory is identical. It corresponds to the second scenario discussed by Lopez-Leon in \cite{lopez2009topological}.

\begin{figure}[b]
\centering
 \includegraphics[width=8cm]{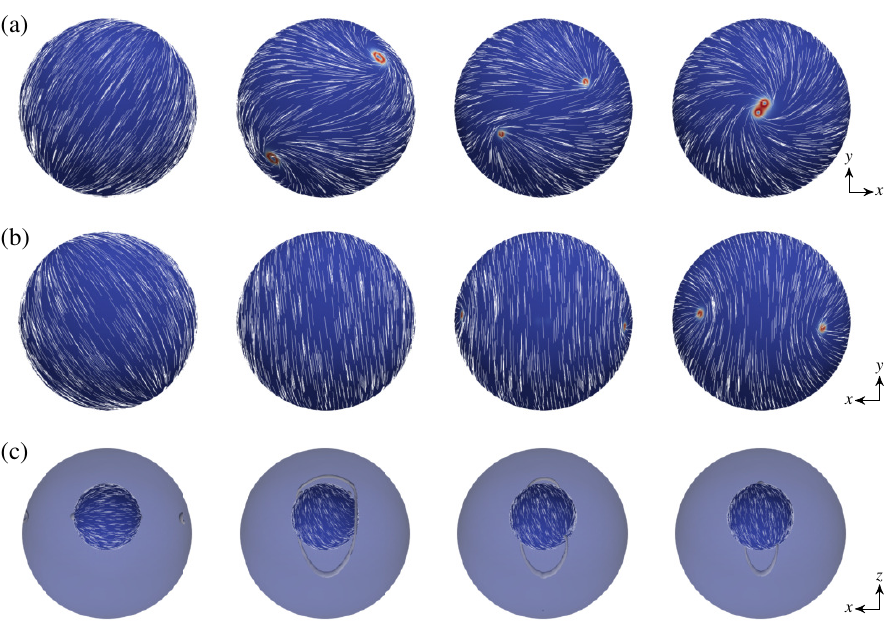}
 \caption{Simulated trajectory of the T$\rightarrow$H transition starting from a +1$^l$, +1$^r$ (left, right) tangential configuration, with $t = -1.79$, $c = 0.2$, $\rho = 0.4$ and $\eta = 4$; we plot \textbf{n} on the inner shell boundary [simulation steps $n$ = 0, 30000, 35000 and 40000 from left to right]. (a) View from the thin part, (b) view from the thick part, and (c) side view. }\label{fgr:2_P_H}
\end{figure}

%\subsubsection{Starting from a tangential configuration with two $+1$point defect pairs; results from simulations}
We are not able to reproduce the 2(+1$^t$) tangential configuration in simulations (see Appendix~\ref{tangentialsim}), hence we compare the experimental observations with simulations from a slightly different initial T3' configuration, with two +1 defect pairs on the sides (Fig~\ref{fgr:2_P_H}). The shell retains the same asymmetry and dimensions as before, with $\eta=4$, and the simulation is done in the same way with a sudden switch from tangential to normal anchoring on the outside interface. 

In the initial stages of the dynamical evolution, the two +1 defects approach and the surrounding director field twists simultaneously, yielding a situation analogous to the experiment in Fig.~\ref{fgr:twodefects}d--e. However, in stark contrast to the experimental trajectory, we then observe a splitting of the two +1 defects into four +\sfrac{1}{2} defects, connected pairwise by U-bent disclinations through the shell. Two +\sfrac{1}{2} defects then merge into a +1$^t$ defect with surrounding twisted director field at the thinnest part of the asymmetric hybrid shell. The remaining two +\sfrac{1}{2} defects migrate to the thickest part of the shell, again without merging. As above, we attribute the discrepancies between simulation and experiment to the differences in shell size and the non-varying LC parameters.

\subsection{Detailed investigation of hybrid shells}\label{hybriddetails}

The experiments in Sections \ref{5cbshelldescript} and \ref{E7-F127-shells} show that the temperature $T_c$ when the gradual change from tangential to normal anchoring has completed is different on the shell in- and outside, as otherwise we would not get the H configuration at intermediate temperatures. But these experiments do not reveal which side has lower $T_c$. In order to elucidate this issue, we conduct a series of reference experiments where we stabilize shells only from one side with F127, the opposite side stabilized by a 1\% SDS solution. It is well established that SDS at this concentration induces strong normal anchoring, hence the SDS-stabilized side is always normal. As the F127-stabilized side is predominantly tangential at room temperature, all shells in the reference experiments start out in H configuration. On heating, the F127-stabilized side gradually changes anchoring, the shell switching to N configuration when $T>T_c$. If $T_c$ is lower on the inside, this happens at a lower temperature than when we reverse the geometry. If $T_c$ is lower on the outside, it will be the other way around.

\begin{figure}
\centering
 \includegraphics[width=\columnwidth]{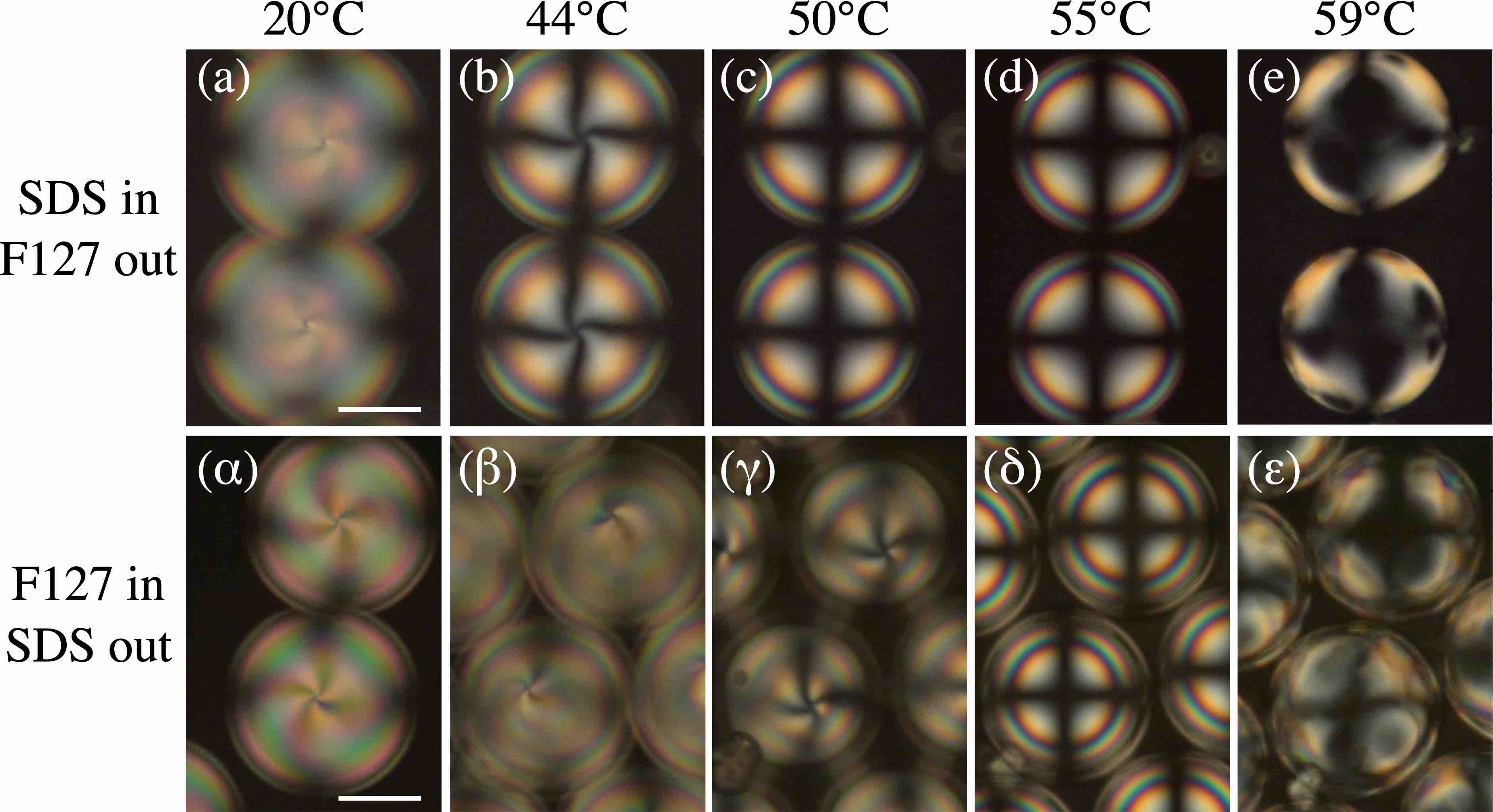}
 \caption{Heating E7 shells stabilized by the normal-aligning surfactant SDS on the inside and F127 on the outside (a--e) or the other way around ($\alpha$--$\epsilon$), from 20$^\circ$C to the clearing point. All shells start out in H configuration, the F127-stabilized side tangential. With F127 on the outside, the shells switch to N configuration above 44$^\circ$C (b--c), whereas this transition is seen only at about 55$^\circ$C when F127 is on the inside ($\gamma$--$\delta$). The tangential-aligning ability of F127 is thus weaker on the outside than on the inside. Scale bars: 50~$\mu$m.}
 \label{fgr:SDS-F127}
\end{figure}

The results are shown in Fig.~\ref{fgr:SDS-F127} and in movies M5 and M6. As predicted, all shells start out in H configuration at 20$^\circ$C, see panels (a/$\alpha$). The texture changes gradually upon heating and at 44$^\circ$C the shells with F127 on the outside and SDS on the inside acquires an N configuration, see Fig.~\ref{fgr:SDS-F127}b. The shells with the inverted stabilizer geometry still have a complex H texture at this temperature (Fig.~\ref{fgr:SDS-F127}$\beta$), turning into N only at 55$^\circ$C (Fig.~\ref{fgr:SDS-F127}$\delta$), 10 degrees higher. The comparison clearly shows that F127 retains its tangential-aligning influence to higher temperatures on the inside than on the outside, and we can thus conclude that all H shells described in sections \ref{5cbshelldescript} and \ref{E7-F127-shells} have the outside close to normal anchoring; the tilt gradually increases, reaching normal alignment at a temperature for which the inside still has tilted (or tangential) alignment.

\begin{figure}[b]
\centering
 \includegraphics[width=8.5cm]{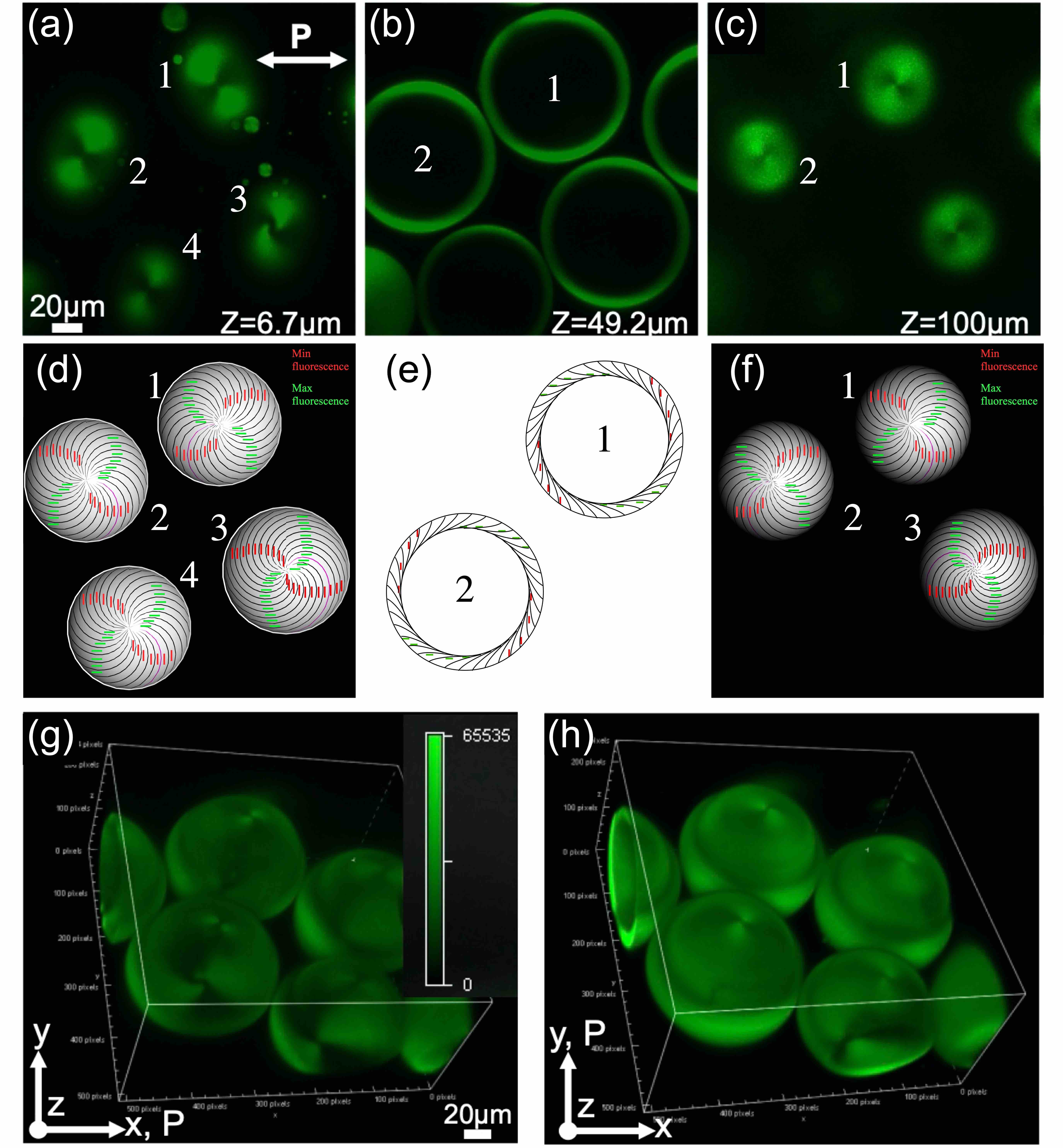}
 \caption{FCPM images of E7 shells in H configuration stabilized by 1\% F127 on in- and outside, taken at 54$^\circ$C. Fluorescence images and corresponding director field sketches (for connecting top to bottom, one field line has been colored pink) are for cross sections near the shell bottom (a/d, $z=6.7$~$\mu$m), mid plane (b/e, $z=49.2$~$\mu$m) and top (c/f, $z=100$~$\mu$m). For clarity, we draw only shells 1--2 in (e). Locations with $\mathbf{n}$ parallel to the exciting laser polarization $\mathbf{P}$ (maximum fluorescence from the BTBP dye) are highlighted with green lines, whereas locations with $\mathbf{n}\perp\mathbf{P}$ (minimum fluorescence) are indicated with red lines. All shells are qualitatively similar to Fig.~\ref{fgr:3steps}g--i, although shells 2--4 have the opposite handedness and the degree of twist varies, with shell 4 having stronger twist than 1--3. (g--h): 3D volume-views of the shells, for $\mathbf{P}$ along $x$ (g) and $y$ (h), respectively. The upper half of each shell appears with artificially reduced thickness due to refraction at the water-LC interface at the shell bottom.}
 \label{fgr:fcpm}
\end{figure}

To get further details with 3D resolution, we investigate four E7 shells in +1$^b$,+1$^t$ H configuration, stabilized by F127 on both sides, by fluorescent confocal polarizing microscopy (FCPM). As seen in the cross-sectional image sequence of Fig.~\ref{fgr:fcpm}a-c and Supplementary Movie M7, the fluorescence emission reveals a twisted director field from bottom to top. If there were no twist, the fluorescence maxima would be along $\mathbf{P}$ and the minima $\perp\mathbf{P}$. We note that the fluorescence patterns at the bottom (Fig.~\ref{fgr:fcpm}a/d) and the top (Fig.~\ref{fgr:fcpm}c/f) are each others' mirror images. This is because the fluorescence intensity reflects the component of the dye molecule's transition moment along the exciting light polarization. This is given by the projection into the horizontal plane of the orientation of the dye molecule, and thus of the director field. As illustrated by comparing panels (d) and (f) in Fig.~\ref{fgr:fcpm}, this projection at the top is the mirror image of that at the bottom. This is also seen in Fig.~\ref{fgr:3steps}g, showing the microscopy perspective of an H configuration shell, with direct comparisons with other perspectives. This means that the apparent projected spiral director field around each +1 defect appears with opposite handedness at the top and bottom for the same shell. 
 
At the shell mid plane (Fig.~\ref{fgr:fcpm}b/e) the projection of ${\bf n}$ into the horizontal plane exhibits a bend from the predominantly tangential inside to the predominantly normal outside. Both interfaces are likely to have tilted alignments; if the outside were perfectly normal, the bend would take place over a rather small distance and we should see strong fluorescence at the top and bottom (of the image, not of the shell) along the shell insides, and on the left and right along the shell outsides, but this is not what we see. The rotated maxima and minima indicate tilted alignments with respect the interface normals.
 %, near $0^\circ$ on the outside {\color{red} inside} and near $90^\circ$ on the inside {\color{red} outside, please check once}, the tilt angle defined with respect to the interface normal. %Especially for shell 2 the sketched director field would reproduce quite well the actual fluorescence, whereas for shell 1 there is a minimum at the top left that is difficult to explain. This may indicate that \textbf{n} on the outside is not uniform across the shell. 
 In Fig.~\ref{fgr:fcpm}g--h, the full 3D FCPM images are shown for two perpendicular orientations of the exciting light polarizations. In (g) the polarization is identical to the cases in (a--c), whereas (h) has the perpendicular polarization.

\subsection{Comparison with PVA as stabilizer}

%As for 5CB shells, the F127-stabilized E7 shells all adopt the 2(+1$^t$) director field when they revert back to tangential configuration at room temperature. The shell thickness does not affect the outcome, see Fig.~\ref{fgr:AfterCooling}. Panel (a) shows shells with similar thickness as those studied above whereas the shells in (b) are several times thicker. We let the shells stand at room temperature for extended time to see if this unexpected single-configuration tangential state is transient or stable. After two days, we see no change: all shells remain in the the 2(+1$^t$) configuration. We do not see a splitting of defects to the competitor 4(+\sfrac{1}{2}$^t$) configuration in any shell. This suggests that the energy barrier for changing configuration is significantly higher than the thermal energy $kT$ at room temperature, thus rendering the observed state metastable after slow cooling from the isotropic phase. 

% \begin{figure}
%\centering
% \includegraphics[width=7cm]{AfterCooling.jpg}
% \caption{Thin (a) and thick (b) E7 shells after heating to isotropic and cooling at moderate rate to room temperature. All shells have 2(+1$^t$) configuration. Scale bar: 100 $\mu$m.}
 %\label{fgr:AfterCooling}
%\end{figure} 

\begin{figure}
\centering
 \includegraphics[width=8.5cm]{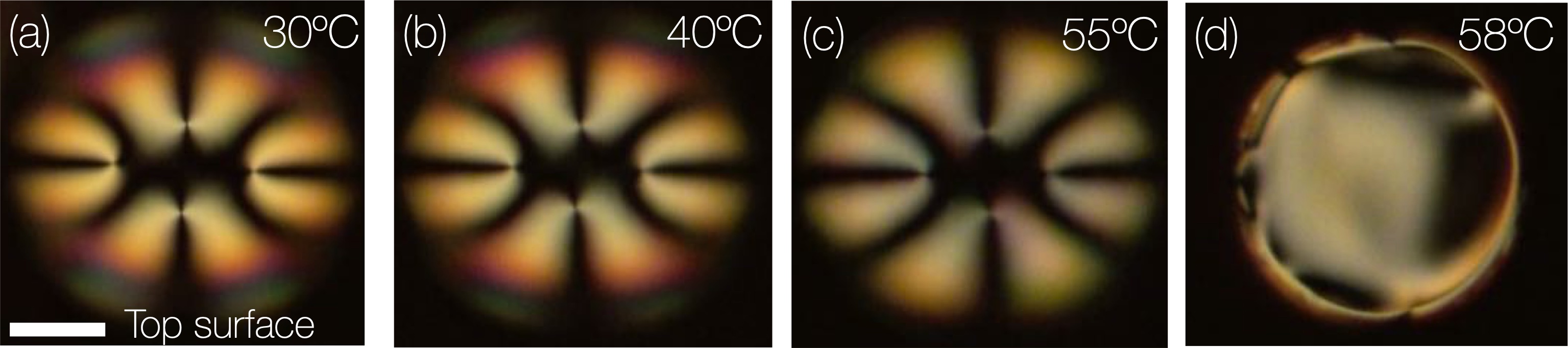}
 \caption{E7 shells stabilized by aqueous PVA solutions on both sides, heated from room temperature to $T_{NI}$. The shells remain in a T3 [4(+\sfrac{1}{2}$^t$)] configuration (a--c) until clearing (d). Scale bar: 50 $\mu$m.}
 \label{fgr:PVA-reference}
\end{figure}

To confirm that the temperature-induced alignment change is indeed characteristic of shells stabilized by the block copolymer F127, we also produce E7 shells stabilized by 1\% aqueous solutions of 85\% hydrolyzed PVA on both sides, heating the shells at 5~K/minute from room temperature to $T_{NI}$, see movie M8 and Fig.~\ref{fgr:PVA-reference}. The T1 configuration texture remains intact constantly throughout the heating process, until clearing starts at $T\approx58^\circ$C, see Fig.~\ref{fgr:PVA-reference}. The only change that can be distinguished prior to clearing is a slight decrease in birefringence.

\section{Discussion}

We summarize all experimental alignment transition temperatures for 5CB and E7 in Table~\ref{tab:redtemps}. In order to compare the two LCs and to relate the transitions to the degree of orientational order rather than to a fix temperature scale, we use an experimental reduced temperature $T_r=(T-T_{NI})/T_{NI}$, with $T$ and $T_{NI}$ in Kelvin. All transitions, regardless of starting configuration, take place at somewhat lower $T_r$ in E7 than in 5CB and all realignment transitions take place at higher $T_r$ when starting from T3 [2(+1$^t$)] than from T1 [$4(+\sfrac{1}{2}^t)$] or T2 [$2(+\sfrac{1}{2}^t),+1^t$]. The latter discrepancy is not surprising, considering that the realignment trajectory starting from T3 is so different from those starting from T1 or T2. The three transition steps are thus not directly comparable between T1/T2 and T3. Nevertheless, for a given LC material, i.e. either 5CB or E7, Step 1 of the T${}^{\phantom{1}\rightarrow}_{\textrm{heat}}$H transition occurs at similar $T_r$ for the T1 and the T3 starting configurations. Note that the non-reduced temperatures are much more different between the two LCs (about 24$^\circ$C for 5CB and about 35$^\circ$C for E7).

\begin{table}
    \centering
    \caption{Experimental reduced alignment transition temperatures $T_r=(T-T_{NI})/T_{NI}$ for 5CB and E7 shells. The three steps of the T$^{~\rightarrow}_{\textrm{heat}}$H
      transition are defined above. The final transition is from hybrid to normal, H$^{~\rightarrow}_{\textrm{heat}}$N.
    } \label{tab:redtemps}
           \begin{tabular}{c|c|c|c|c|c|c|c}
        LC & Start. config. & $T_r^{Step 1}$ & $T_r^{Step 2}$ & $T_r^{Step 3}$ & $T_r^{H\rightarrow N}$\\\hline
        5CB & T1 & -0.040 & -0.013  & -0.006 & -0.001 \\
        \textbf{E7} & \textbf{T1} &  $\mathbf{-0.08}$ & $\mathbf{-0.05}$ & $\mathbf{-0.03}$ & $\mathbf{-0.005}$ \\
        5CB & T2 & - & -0.034  & -0.0058 & -0.001 \\
        \textbf{E7} & \textbf{T2} & - & $\mathbf{-0.081}$  & $\mathbf{-0.037}$ & $\mathbf{-0.005}$ \\
        5CB & T3 & -0.034 & -0.002 & - & -0.001 \\
        \textbf{E7} & \textbf{T3} & $\mathbf{-0.072}$ & $\mathbf{-0.027}$  & $\mathbf{-0.013}$ & $\mathbf{-0.006}$\\  
     \end{tabular}
  
\end{table}

For both LCs and all starting configurations, there is significant variation in $T_r$ of Step 2, after which a half-shell director field twist arises. As for Step 3, on the other hand, after which the entire shell adopts a twisted +1$^b$,+1$^t$ H configuration, we find similar $T_r$ for starting configurations T1 and T2, for a given LC material. The transition to normal (H${}^{\phantom{1}\rightarrow}_{\textrm{heat}}$N) takes place very near the clearing point ($T_r=0$) in all shells, at a $T_r$ that is, within experimental error, independent of the starting configuration. Measured in normal temperature scale, the H${}^{\phantom{1}\rightarrow}_{\textrm{heat}}$N transition takes place at very different values, about 35.2$^\circ$C for 5CB and about 57.3$^\circ$C for E7.

\subsection{Why is the alignment tangential far below $T_{NI}$ but normal near $T_{NI}$?}\label{whyalignmentchange}
While it is experimentally challenging to distinguish tangential from slightly tilted anchoring, our assumption is that both shell interfaces are initially tangentially aligned at room temperature, thus giving us true T shells to start with. This assumption is based on the fact that we see no qualitative differences between the room temperature shell textures of shells stabilized by F127 and by PVA (compare Fig.~\ref{fgr:fourdefects}a and Fig.~\ref{fgr:PVA-reference}a) respectively, the latter of which is well known to give strong tangential anchoring. Most significantly, the prevalence of the T1 and T2 configurations at room temperature, exhibiting two or four +\sfrac{1}{2} disclinations, respectively, shows that the alignment on both in- and outside is tangential, at least in the vicinity of the disclinations. 

We see textural changes upon heating and assume, based on the experiments in Fig.~\ref{fgr:SDS-F127}, that the alignment on the shell outside tilts away from the tangential conditions before the shell inside, and the tilt magnitude increases with increasing temperature. If tangential anchoring is lost everywhere on the shell outside, it can no longer support +\sfrac{1}{2} disclinations. Hence the $\pi$-lines discussed in  Figures \ref{fgr:fourdefects} and \ref{fgr:threedefects}, connecting pairs of +\sfrac{1}{2} defects, must then coincide with the U-turned disclinations within the shell ( as observed in simulations). An alternative may be that the outside remains tangentially aligned locally around the +\sfrac{1}{2} defects and the tilt appears only in the defect-free regions. The shell inside remains tangentially aligned for a longer length of time, but eventually allows for tilted alignment %(again, in case +\sfrac{1}{2} disclinations are present, the fully tangential alignment must prevail at least locally around the defects) 
and then normal alignment, phase-shifted compared to the shell outside.

The experiments in this paper, together with those reported previously for 8CB in \cite{liang2011liquid}, show that this gradual realignment is not a function solely of absolute temperature, but it depends strongly on the LC. While $T_r$ for a certain transition listed in Table~\ref{tab:redtemps} is not identical for 5CB and E7 shells, it is always of the same order of magnitude. It is similar enough for the transition to be driven primarily by a change in the LC. A secondary, much weaker, influence may be linked to temperature dependant behavior of the surrounding F127 solution, in particular regarding the degree of hydration of its different blocks (see below), explaining why a certain $T_r$ is always lower for E7 (high $T_{NI}$) than for 5CB (low $T_{NI}$). 

F127 is an amphiphilic block copolymer. In contrast to low molar mass amphiphiles, which give normal alignment due to their radial alignent at the shell--water interface unless the concentration is very low \cite{Sharma2018influence}, the impact of a block copolymer is more ambiguous. %This is because both blocks will be in contact with the liquid crystal, as dictated by the sheer size of the molecule. 
F127 has two hydrophilic PEO blocks, which are well hydrated near room temperature \cite{Alexandridis1994,Linse1993a,Linse1993}, thus bringing water in contact with the shell. This promotes tangential alignment \cite{Volovik.1983, Brake2002}. In contrast, the hydrophobic PPO block at the center of the molecule is not hydrated. Its aliphatic nature promotes contact with the alkyl chain of LC molecules, favoring normal alignment. Because all three F127 blocks will be in contact with the LC shell, the overall influence on \textbf{n} at the LC boundaries is weak; two blocks favor tangential alignment, but they also prefer the water phase to the LC, and one block, preferring the LC, favors normal alignment. F127 is thus immensely interesting as it provides good shell stability through its adsorption at the LC--water interface, yet it gives little preference between tangential and normal alignment.

The anchoring of \textbf{n} at the shell boundary is dictated by the interfacial free energy density $F_S$ (see equation (9) and (12)), i.e. $F_S= W \left( \Qvec - \Qvec^{\perp} \right)^2$, where $\Qvec^{\perp}$ is the tangential projection on the plane of the boundary and $W$ is a measure of the anchoring strength. In particular, if $W>0$, then tangential alignment is favoured so that $\Qvec = \Qvec^{\perp}$ whereas $W<0$ favours normal alignment.
%is a baseline interfacial tension independent of the director orientation, \textbf{c} is the boundary normal, \textbf{n} is the director, and $W$ is a dimensionless parameter called the anchoring strength. If $W>0$, tangential alignment at the interface is favored ($(\mathbf{c}\cdot\mathbf{n})^2=0$ gives the minimum interfacial energy, equal to $\gamma$) whereas $W<0$ favors normal alignment ($(\mathbf{c}\cdot\mathbf{n})^2=1$ gives the maximum interfacial energy reduction, reducing it to $\gamma-W$). By virtue of its relation to the orientational order of the liquid crystal $W$ must depend on the orientational order parameter $S$ of the liquid crystal. Thus, as we approach the clearing point, the impact of $W$ should continuously decrease. 
We conjecture that, for F127, $W>0$ but with low magnitude, proportional to $S_b$, where $S_b$ is a bulk scalar order parameter that can be computed from the minimizers of the bulk energy in Equation (12). The slight emphasis for tangential alignment can be understood by considering the two large PEO blocks compared to the single smaller PPO block. Far below $T_{NI}$, where $S_b\approx0.7$, $W$ is large enough to impose tangential alignment despite the elastic energy penalty imposed by the resulting topological defects. However, as $T$ approaches $T_{NI}$ on heating, $S_b$ decreases significantly and therefore $W$ decreases. The energetic penalty of not adopting tangential alignment on the outer shell surface decreases compared to the elastic energy cost of a tangential configuration with defects. This would qualitatively explain why the alignment gradually tilts over from tangential at room temperature to normal near $T_{NI}$.  

%On heating, the hydrophilic blocks of F127 get less hydrated \cite{Alexandridis1994,Linse1993a,Linse1993}, hence the tangential influence is reduced, i.e. $W$ decreases. This may explain why the reduced temperatures of all alignment transitions are lower for E7 than for 5CB: because $T_{NI}$ is higher for E7 than for 5CB, a certain value of $S_b$ occurs at higher temperature for E7, thus at a lower $W$.  However, the temperature dependence of F127 PEO block hydration is only a secondary effect, the temperature dependence of the LC order parameter clearly dominating the behavior.

\begin{figure}
\centering
 \includegraphics[width=8.5cm]{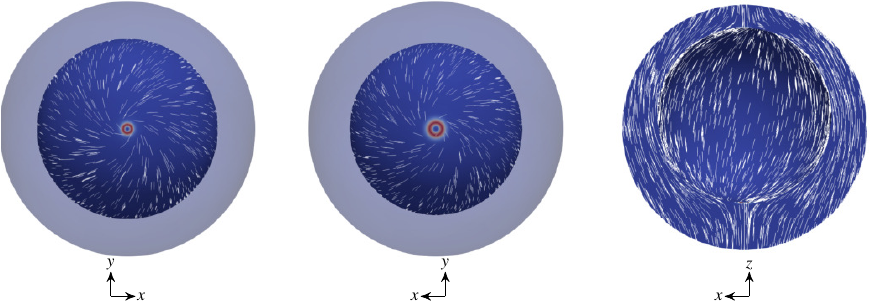}
 \caption{Hybrid shell obtained with weak tangential anchoring on the outside ($W = 10^{-10} \mathrm{Jm}^{-2}$) and strong tangential anchoring on the inside ($W = 10^{-2} \mathrm{Jm}^{-2}$), viewed from the thinnest part (top), thickest part (bottom) and side, respectively, from left to right. A twisted +1$^b$,+1$^t$ director field arises, similar to Fig.~\ref{fgr:hybrid}a, with normal outer anchoring.} 
 \label{fgr:hybrid2}
\end{figure}

To test if this conjecture is plausible, we numerically simulate the director field of a shell with strong tangential anchoring on the inside but very weak tangential anchoring on the outside ($W = 10^{-10} \mathrm{Jm}^{-2} $), see Fig.~\ref{fgr:hybrid2}. Indeed, this results in an H configuration essentially identical to that in Fig.~\ref{fgr:hybrid}a, obtained with strict normal anchoring on the outside. %This demonstrates that the configuration of an LC shell is indeed not solely determined by the boundary conditions, but the energy cost of elastic deformation within the shell can have a significant impact, as seems to be revealed in the F127-stabilized shells as the temperature is raised.

Why does the realignment process start and finish earlier on the outside than on the inside? We propose that the negative curvature on the shell inside forces the LC to be more in contact with the hydrophilic blocks on the inside than on the outside. While for small molecules the curvature of a 100~$\mu$m diameter shell may be considered negligible, this is not necessarily the case for molecules on the scale of a block copolymer. On the shell outside, with positive curvature, the LC around a point of contact with a PPO block curves \textit{away} from the surrounding PEO blocks. They can thus avoid LC contact reasonably well, easily fluctuating outwards into the continuous aqueous phase, improving their hydration and increasing the entropy. This increases the packing density of PPO blocks on the LC shell outside, weakening the tangential-aligning influence of PEO blocks. 

On the inside, in contrast, the LC around a point of contact with a PPO block curves \textit{towards} the surrounding PEO blocks. The hydrated PEO blocks are thus more likely to be in contact with the LC on the inside than on the outside, strengthening the tangential-aligning influence there. Our conjecture needs to be corroborated and we hope our results can inspire thrusts with molecular dynamics simulations. 

An alternative explanation has been proposed by Durey and Lopez-Leon in a reprint that we were recently made aware of \cite{durey2018temperature}, and previously also in a study by Lopez-Leon and co-workers focusing on cholesteric shells \cite{tran2017change}. They report a similar change in alignment as described above, for PVA-stabilized shells very near $T_{NI}$, on heating at 0.01~K/min. We do not see this in our reference experiments with PVA (Fig.~\ref{fgr:PVA-reference} and Movie M8), probably due to faster heating. In \cite{durey2018temperature} the authors propose that a transition to normal anchoring is driven by a thin surface layer of LC material that turns isotropic. We find this explanation unlikely, however, as the isotropic--nematic interface in cyanobiphenyl-based mixtures, with co-existing isotropic and nematic phases, has been shown to induce tangential alignment \cite{kim2015dielectrophoretic}.

We do not know the degree of hydrolysis of the PVA used in \cite{durey2018temperature}, but as it is typical to use 85\% hydrolyzed PVA for stabilizing LC shells (fully hydrolyzed PVA tends to give poor interface stability), we assume that this may have been the case. The remaining 15\% are then polyvinylacetate which is less hydrated than polyvinylalcohol. While it is not a block copolymer, there is a variation in hydration between sections also in this stabilizer, and this could, qualitatively, give rise to the same features as discussed above. However, compared to F127 the tangential-aligning influence of PVA, even if not fully hydrolyzed, is much stronger, explaining why Durey and Lopez-Leon observed realignment only over tenths of degrees from the clearing point, whereas we see it over tens of degrees with F127-stabilized shells.  

\subsection{Why do we see twisted director fields around +1 point defects in hybrid configurations}

A common feature of the experiments and simulations are twisted director fields around the +1 point defects on the tangential inner surface, for hybrid configurations. We believe that the twist is a consequence of the elastic anisotropy, i.e., when the splay elastic constant is larger in magnitude than the twist elastic constant. The elastic anisotropy is captured by the parameter $\eta = 2 \left(\frac{K_1}{K_2} - 1\right)$. Our simulations show pure splay director fields around the +1 point defects for $\eta =0$ and the twist naturally appears on increasing $\eta$. The modelling at this stage does not capture the temperature dependence or even the material-dependence of the elastic constants. This will be pursued more systematically in future work.

The first +1 point defect to form in a hybrid configuration always localises at the thinnest point of the shell, in both experiments and simulations. For topological reasons,  we must have a total topological charge of +2 on the tangential inner surface. In terms of the modelling, we find two competing hybrid configurations, $+1^t, +1^b$ and $+1^t, 2(+\sfrac{1}{2}^b)$, both of which are local energy minimizers. In the experiments, we only observe the $+1^t, +1^b$ hybrid configuration. The results are sensitive to a number of geometrical, material and environmental factors and since the modelling is largely phenomenological, we do not expect perfect agreement with experiments at this stage. We speculate that $+1^t, 2(+\sfrac{1}{2}^b)$ will lose its stability when the radius of the inner sphere becomes large enough, the thickness of the thick part becomes smaller, or $\eta$ is larger. This will be investigated in future work.

% {\color{blue} We do not observe twisted director fields around the +1/2 disclination lines near the bottom, which are well separated in the $+1^t, 2(+1/2^b)$ hybrid configuration. We speculate that a twisted director field around the end points of the two well-separated +1/2 disclinations would induce a more elastically distorted configuration throughout the shell than an untwisted director field, for moderate values of $\eta$.} % However for values of $\eta$ larger than the ones considered in this manuscript, we may observe twisted director fields around the +1/2 defects too. 

\subsection{What determines the tangential director field?}\label{finalconfigurationexplanation}
The simulations reveal (Appendix~\ref{tangentialsim}) that the energy-minimizing tangential configuration is T1 [4(+\sfrac{1}{2}$^t$)], yet not all of the freshly prepared shells are in this configuration, some 15\% showing the T2 [2(+\sfrac{1}{2}$^t$),+1$^t$] configuration and about 10\% the T3 [2(+1$^t$)] configuration. In contrast, when the shells are cooled relatively slowly, going through the full sequence of realignment, all shells adopt the T3 configuration. Additional experiments (not shown) demonstrate that this is independent of shell thickness, and that it is stable in time (no change within 2 days confirmed).

We propose that each of the T configurations corresponds to a local energy minimum, with energy barriers for change from one to the other that are significantly greater than thermal energy at room temperature. When the shells are cooled slowly, the route via the +1$^b$,+1$^t$ hybrid state templates a T state with two +1 defects, which move up to the thinnest point of the shell to form a 2(+1$^t$) T3 configuration once the tangential state is fully developed. Moreover, as long as a tilted alignment prevails across an interface, no +\sfrac{1}{2} disclinations can form at that interface, hence only the T3 configuration is compatible with a shell that may appear like a true T shell but actually has a very weakly tilted interface alignment, away from tangential across either or both interfaces.

However, if the shells are quench cooled very rapidly, as is the case at the end of the shell production, the isotropic--nematic phase transition takes place during strongly non-equilibrium conditions, and a random distribution of all possible T configurations is seen. The isotropic state may be temporarily supercooled to near room temperature, ensuring that the stable interface alignment is truly tangential once the nematic order develops, allowing +\sfrac{1}{2} disclinations to form. This would explain why we see all T configurations in pristine shells. In fact, the simulations at constant LC parameters may in some respects be closer to these experimental conditions. Again the energy barriers for changing from one configuration to the other prevents a shell with a certain configuration to switch to another configuration, even if the other configuration has lower free energy.

%Video~\ref{vid:PRSTPER.4.010101} 
%illustrates several features new with REV\TeX4.2,
%starting with the \texttt{video} environment, which is in the same category with
%\texttt{figure} and \texttt{table}.%
%\begin{video}
%\href{http://prst-per.aps.org/multimedia/PRSTPER/v4/i1/e010101/e010101_vid1a.mpg}{\includegraphics{vid_1a}}%
% \quad
%\href{http://prst-per.aps.org/multimedia/PRSTPER/v4/i1/e010101/e010101_vid1b.mpg}{\includegraphics{vid_1b}}
% \setfloatlink{http://link.aps.org/multimedia/PRSTPER/v4/i1/e010101}%
% \caption{\label{vid:PRSTPER.4.010101}%
%  Students explain their initial idea about Newton's third law to a teaching assistant. 
%  Clip (a): same force.
%  Clip (b): move backwards.
% }%
%\end{video}
%The \verb+\setfloatlink+ command causes the title of the video to be a hyperlink to the indicated URL; it may be used with any environment that takes the \verb+\caption+ command. The \verb+\href+ command has the same significance as it does in the context of the \texttt{hyperref} package: the second argument is a piece of text to be  typeset in your document; the first is its hyperlink, a URL.

%\textit{Physical Review} style requires that the initial citation of figures or tables be in numerical order in text, so don't cite Fig.~\ref{fig:wide} until Fig.~\ref{fig:epsart} has been cited.

\section{Conclusions and outlook}
We have undertaken a comprehensive study of nematic shells stabilized by the amphiphilic block copolymer F127, using two different LC materials: 5CB and E7. Our most notable findings focus on the dynamic tuning of the boundary conditions for the shells, using temperature as the control variable. We experimentally record highly informative director field reconfigurations from tangential to hybrid and then to normal, before a transition to the isotropic phase, through carefully controlled heating experiments. Subsequent cooling demonstrates the reverse sequence. The experiments illustrate different combinations of topological defects and their trajectories during the reconfigurations, all of which can be used for novel control strategies for materials design. 

We provide original explanations for the fact that F127 can promote both tangential and normal anchoring, tangential preferred at low temperature. We also note that the inner surface gives stronger tangential anchoring as a consequence of the interplay of curvature and the chemical composition and size of F127. We complement our experiments with modelling and simulations, which capture key experimental details and demonstrate the enormous complexity of the solution landscapes in nematic shells. Our modelling includes hybrid shells for the first time. 

Our work is a significant step forward in the design and understanding of LC shells, and it demonstrates the power of block copolymers---so far largely ignored by the community---as LC shell stabilizers. The dynamic tuneability of topology, with varying number of defects distributed in very different ways across the shell, or indeed no defects at all, opens for interesting sensors, not only of temperature but of any stimulus that affects the nematic order parameter and/or the behavior of the block copolymers at the curved LC--water interfaces. The comparatively large LC shell could act as a powerful amplifier of events taking place at the molecular scale \cite{Carlton2013}, with no need for POM interrogation if selectively reflecting cholesteric LCs are used. Equally interesting is the opportunity to easily tune the programmed shape change response of shell-shaped LCE actuators \cite{fleischmann2012one,jampani2018micrometer,jampani2019liquid}. For flat LCE films there are many impressive demonstrations of how targeted positioning of topological defects can be used to radically change the shape shifting during actuation \cite{white2015programmable}, but such control has so far been difficult to achieve in LCE shells. If block copolymer stabilizers offer the same tuneability for LCE precursor materials as for cyanobiphenyls, this obstacle can be removed.

\begin{acknowledgments}
We gratefully acknowledge financial support from the European Research Council under the European Union’s Seventh Framework Programme (FP/2007-2013)/ERC (JPFL, consolidator project INTERACT, grant ID. 648763), the Fonds National de la recherche Luxembourg (JN, grant ID 6992111 and VSRJ, grant ID C17/MS/11703329/trendsetter) and the German Academic Exchange Service (DAAD) and the IRTG (H-LL). AM is supported by a Leverhulme International Academic Fellowship and the University of Strathclyde's New Professors Fund. AM gratefully acknowledges a Visiting Professorship at the University of Bath and an OCIAM Visiting Fellowship at the University of Oxford. YW, AM and JL acknowledge support from ICERM in Brown University, December 2019.
\end{acknowledgments}

\appendix

\section{Explanation of apparent oval shape of tangential-aligned shells}\label{ApxPOM}

A curious effect is that the projection of the fully tangential-aligned shells as observed in POM appears oval rather than circular, although interfacial tension certainly requires the shells to be spherical. This is an optical illusion that can be explained by the details of the director field. With the four defects located near the shell top in Fig.~\ref{fgr:fourdefects}a, \textbf{n} mainly splays outside defects 1 and 2, whereas it primarily bends outside 3 and 4. Due to the spherical geometry of the shell, the effective birefringence $\Delta n_{\textrm{eff}}$ is continuously reduced towards the perimeter in the region of director splay. This is because, to an observer at the microscope, the optic axis, equivalent to \textbf{n}, in this region tilts away from the sample plane, increasingly so the further we are from the center of the curved surface, hence $\Delta n_{\textrm{eff}}$ decreases towards the perimeter. This reduction in $\Delta n_{\textrm{eff}}$ roughly compensates for the increase towards the perimeter in the optical path length of light passing vertically through the shell. Together, the two effects result in an interference color that stays almost the same from center to edge, here in the bright yellow order of the Michel-L\'evy chart. 

In contrast, where \textbf{n} bends outside the 3--4 defects, the optic axis is oriented such that it retains its full length despite the presence of curvature, hence the full $\Delta n$ of the LC is experienced. As the increase in optical path length towards the shell perimeter now takes place at constant $\Delta n$, the effective interference moves to the right in the Michel-L\'evy chart, to a more saturated pink-blue regime. The shell then appears darker, giving the false impression that its extension in the 3---4 direction is less than that in the 1---2 direction. Being aware of this feature can help to map out the director field distinctly, even without the use of a first-order $\lambda$ plate. 

As is clear from the decreasing apparent overall size of the shells as they are heated from tangential via hybrid to normal configuration, the change in effective refractive index also affects the lensing effect that the curved LC gives rise to. Because this is further complicated by density variations with temperature, as well as the effect of disclinations, potentially affecting the thickness at different points of the shell, a complete analysis of this aspect is outside the scope of this paper.

\section{Numerical Methods}\label{ApxNum}

Here we give a detailed description of our numerical methods. The experimentally observable states are modelled by the locally stable points, i.e., local or global minimizers, of the LdG free energy.  We point out that there are typically multiple local energy minimizers for highly nonlinear and non-convex problems such as the LdG minimization problem and whilst the global minimizer may be most frequently observed, local minimizers also have a basin of attraction and stability \cite{Yin2020prl}. In particular, in an experiment, a local minimizer will not relax to the global minimizer with minimum free energy because all stable states are separated by an energy barrier.

 % we use spectral methods to compute locally stable critical points or local minimizers of the LdG free energy, for a given set of values of $\left(t,\eta,\xi_R,\omega_i\right)$, which model experimentally observable states.
 
 In order to compute these locally stable points, we use spectral methods to discretize the order parameter $\Qvec$. Spectral methods are efficient numerical methods with high accuracy \cite{shen2011spectral}. Several previous studies have shown that the spectral method is a powerful tool to numerical study LdG free energy \cite{hu2016disclination, tong2017defects, wang2017topological,wang2018formation,canevari2020well}. The key idea to apply the spectral method to our system is to use a bispherical polar coordinate system 
$(\xi, \mu, \varphi)$ \cite{wang2017topological}, which is given by 
\begin{equation}
\rho = \frac{a \sin \mu}{\cosh \xi - \cos \mu}, \quad z = \frac{a \sinh \xi}{\cosh \xi - \cos \mu},  \quad \phi = \varphi.
\end{equation}
where $(\rho, z, \phi)$ are standard cylindrical coordinates,
\begin{equation}
 a =  \frac{1}{2c} \sqrt{(1 - \rho^2 - c^2)^2  - 4 c^2 \rho^2)};
\end{equation}
for given $c$ and $\rho$. Letting  $\zeta = 2 (\xi - \xi_0)/(\xi_1 - \xi_0) - 1$, we map the original domain to 
$$\Omega = \{(\zeta, \mu, \varphi) | -1 \leq \zeta \leq 1, 0 \leq \mu < \pi, 0 \leq \varphi < 2\pi \}.$$
Then, we can expand the tensor function $\Qvec({\bf r})$ in terms of special functions: real spherical harmonics of $(\mu, \varphi)$ and Legendre polynomials of $\zeta$ \cite{wang2017topological}. We write
\begin{equation}\label{expand}
  \Qvec{(\bf r)} = \sum_{l = 0}^{ L - 1}\sum_{m = 1 - M}^{M - 1} \sum_{n = |m|}^{N-1} {\bf q}_{lnm} P_l(\zeta) Y_{nm}(\mu, \varphi).
\end{equation}
where $L, N, M$ specify the truncation limits of the expanded series, $Y_{nm}$ is defined by
\begin{equation}
Y_{nm} = P_{n}^{|m|}(\cos \mu) X_m(\varphi),
\end{equation}
where $P_n^m$ $(m \geq 0)$ are the normalized associated Legendre polynomials and $X_m$ is given by
\begin{equation}
X_{m}(\varphi) =
\begin{cases}
\cos m \varphi \quad m \geq 0, \\
\sin |m| \varphi \quad m < 0.
\end{cases}
\end{equation}
Due to the original symmetry of the $\Qvec$ tensor, only 5 elements of ${\bf q}_{lnm}$ are independent. Inserting (\ref{expand}) into (\ref{LdG_dimless}), we obtain a free energy as a function of these unknown tensor order parameter elements, ${\bf q}_{lnm} $, denoted by $F({\bf q})$.
The re-defined free energy function is then minimized by using a standard optimization method, such as L-BFGS \cite{wright1999numerical} that treats the independent elements of tensor ${\bf q}_{lnm} $ as variables. Most of simulations results in this paper are obtained by taking $(N, L, M) = (64, 16, 64)$.

We can track the dynamics to equilibria by using the gradient flow algorithm, based on the principle that systems evolve to a local energy minimum with time or equivalently, free energy decreases with time till we settle into a local energy minimum.
A general gradient flow equations can be obtained from a prescribed energy-dissipation law \cite{Giga2017}
\begin{equation}\label{ED_Q}
\frac{\dd}{\dd t} \mathcal{F}[\widetilde{\Qvec}] = - \int_{\Omega} M(\x) \widetilde{\Qvec}_t \cdot \widetilde{\Qvec}_t \dd t,
\end{equation}
where $M(\x)$ is a positive-definite matrix that determines the dynamics approaching the equilibrium, $\widetilde{\Qvec} \in \mathbb{R}^5$ is the vectorized order parameter $\Qvec$. By a discrete energetic variational approach \cite{liu2019lagrangian}, the gradient flow equation of $\widetilde{\Qvec}$ corresponds to energy-dissipation law (\ref{ED_Q}) can be approximated by the gradient flow equations of
the coefficients ${\bf q}_{lnm}$, given by 
\begin{equation}\label{GD_q}
\begin{aligned}
& \frac{\dd}{\dd t} q_{lnm}^i = - \frac{\pp F({\bf q})}{\pp q_{lnm}^i}, \quad i = 1, \ldots 5, \\
% & \\
\end{aligned}
\end{equation}
with a proper choice of $M(\x)$. 
We can solve (\ref{GD_q}) by an explicit Euler method with adaptive stepsize, which updates the $q_{lnm}^i$’s according to
\begin{equation}
\begin{aligned}
& \left(  q_{lnm}^i \right)^{n+1} = \left(  q_{lnm}^i \right)^{n} - \epsilon_n \frac{\pp F({\bf q})}{\pp q_{lnm}^i} \Big|_{{\bf q} = {\bf q}^n}, \\
% & \\
\end{aligned}
\end{equation}
where $\epsilon_n$ is the step length for n-th iteration. We choose $\epsilon_n$ as Barzilai-Borwein step size \cite{barzilai1988two}.
We use the gradient flow algorithm to mimic transitions from fully tangential shells to hybrid shells and vice-versa in what follows.

\section{Simulated fully tangential shells}\label{tangentialsim}

\begin{figure}
\centering
 \includegraphics[width=8.5cm]{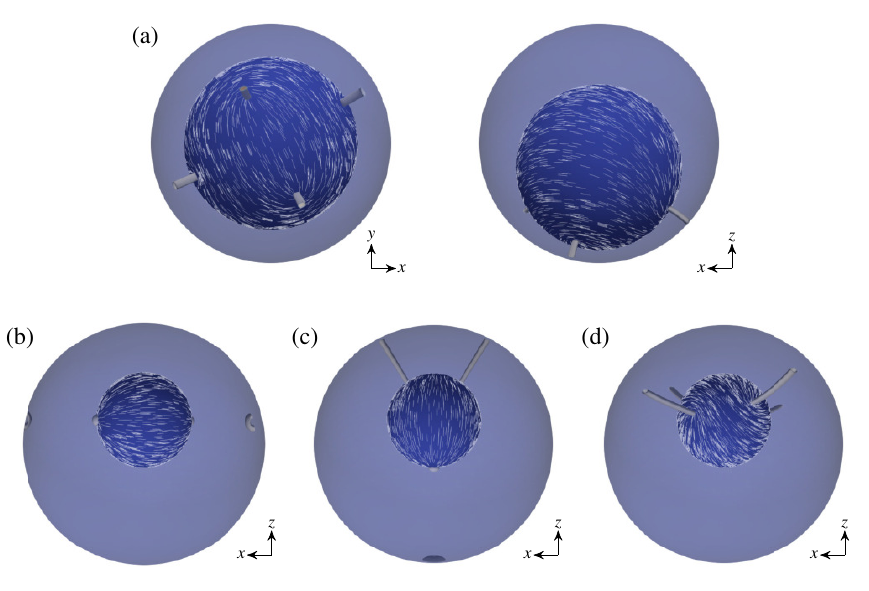}
 \caption{Free energy minimization of thin shells in T configuration (a) suggests that energy minimizers have four +\sfrac{1}{2} disclinations near the thinnest point (4(+\sfrac{1}{2}$^t$)). The parameters are $\xi_R = 50$, $R\approx1$~$\mu$m, $c = 0.2$ and $\rho = 0.7$, with $\eta = 0$ ($K_1=K_2=K_3$). %In (a), the reduced temperature $t = -1.79$, and in b, $t = 0.75$. 
 The shell is viewed from the thin side [left] and from the side [right] respectively. For thick shells (b--d, viewed from the side; $c = 0.2$ and $\rho=0.4$, all other parameters as above), all three possible defect combinations are found, 2(+1) in (c), +1,2(+\sfrac{1}{2}) in (d) and 4(+\sfrac{1}{2}) in (e). We numerically capture the singular regions, drawn with tubes around each defect. In all cases, the depicted director field reflects the internal shell boundary.} 
 \label{fgr:simtangential}
\end{figure}

To benchmark our numerics with respect to previous simulations, comparing also with our experimental results, we simulate T configurations using the one-constant approximation, i.e., $\eta = 0$. For thin shells, only the T3 [4(+\sfrac{1}{2}$^t$)] configuration is found by direct minimization, see Fig.~\ref{fgr:simtangential}(a). For thicker shells, we can find three equilibrium T configurations also in simulation, with defects no longer concentrated to the thinnest point: a +1$^l$,+1$^r$, where $l$ stands for 'left
 and $r$ for 'right' (Fig.~\ref{fgr:simtangential}b), a +1$^{b}$,2(+\sfrac{1}{2}$^{t}$) (Fig.~\ref{fgr:simtangential}c) and a 4(+\sfrac{1}{2}$^t$) arrangement (Fig.~\ref{fgr:simtangential}d). The simulation results are largely compatible with previous reports on simulations of shells in T configuration \cite{lopez2011frustrated,sevc2012defect,koning2013bivalent,koning2016spherical}.

%\nocite{*}

%\bibliographystyle{apsrev}
\bibliographystyle{apsrev}
\bibliography{Bibliography.bib}% Produces the bibliography via BibTeX.

\end{document}